\documentclass[fleqn]{WileyMSP-template}

\pdfoutput=1

\usepackage[utf8]{inputenc}
\usepackage{amsmath}
\usepackage{amsfonts}
\usepackage{amssymb}
\usepackage{graphicx}
\usepackage{color}
\usepackage{bm}
\usepackage{helvet}
\usepackage{placeins}
\usepackage{float}
\usepackage{xspace}
\usepackage{microtype}
\usepackage{caption}
\usepackage{mathrsfs}

\newcommand{\bra}[1]{\langle#1|}
\newcommand{\ket}[1]{|#1\rangle}
\newcommand{\BK}{{\bm{k}}}

\newcommand{\C}[1]{\hat{c}^{\phantom{\dagger}}_{#1}}
\newcommand{\CD}[1]{\hat{c}^{\dagger}_{#1}}
\newcommand{\N}[1]{\hat{n}^{\phantom{\dagger}}_{#1}}
\newcommand{\RE}{\text{Re}}

\newcommand{\DawsonF}{\text{D}_+}
\def\intlimits{\int}

\newcommand{\D}{\ensuremath{\text{d}}}
\newcommand{\IMI}{\text{i}}
\newcommand{\EXP}{\text{e}}
\newcommand\mybar[1]{\bar{#1}}

\newcommand\tIni{t_{\text{ini}}}
\newcommand\tFin{t_{\text{fin}}}

\begin{document}\justifying

  \pagestyle{fancy}
  \lhead{\thepage}
  \rhead{}%

  \title{\sloppy\mbox{Photoinduced} \mbox{Prethermalization} \mbox{Phenomena} \mbox{in} \mbox{Correlated} \mbox{Metals}}
  
  \maketitle
  
  \author{Marc Alexander}, \author{Marcus Kollar}
  
  \dedication{}
  
  \begin{affiliations}
    Theoretical Physics III, Center for Electronic
    Correlations and Magnetism, Institute of Physics, University of Augsburg, 86135 Augsburg, Germany\\
  \end{affiliations}
  
  \keywords{Pump-probe spectroscopy, prethermalization, optical conductivity}
  
  \begin{abstract}
    We study prethermalization phenomena in weakly interacting Hubbard
    systems after electric-field pump pulses with finite duration.  We
    treat the Hubbard interaction $U$ up to second order, applying the
    prethermalization paradigm for time-dependent interaction
    protocols, and the electric field strength beyond linear order.  A
    scaling behavior with pulse duration is observed for the absorbed
    energy as well as individual prethermalized momentum occupation
    numbers, which we attribute to the leading quadratic orders in
    interaction and electric field.  We show that a pronounced
    non-thermal momentum distribution can be created with pump pulses
    of suitable resonance frequencies, and discuss how to distinguish
    them from thermal states.
  \end{abstract}

  \section{Introduction}

  \subsection{Photoexcitation of correlated electrons}

  Using pump-prope spectroscopy it is possible to observe the
  excitation and relaxation of interacting electron systems in real
  time, while potentially creating states or phases that do not occur
  in
  equilibrium~\cite{basov_towards_2017,wang_theoretical_2018,de_la_torre_colloquium_2021}. Typically
  three stages are involved in this
  procedure~\cite{aoki_nonequilibrium_2014}: an initial laser pulse
  which excites the electronic system, followed by its relaxation due
  to the scattering of electrons, and finally a transfer of energy to
  lattice degrees of freedom~\cite{yonemitsu_theory_2006}. The
  photoexcited state may involve nonequilibrium steady states such as
  periodically driven Floquet
  states~\cite{wang_observation_2013,bukov_universal_2015,canovi_stroboscopic_2016,herrmann_floquet_2017,hubener_creating_2017,tindall_analytical_2021},
  the dynamical generation of interactions~\cite{mentink_ultrafast_2015,mikhaylovskiy_ultrafast_2015},
  or states at effectively negative temperature~\cite{braun_negative_2013}. Its
  relaxation may pass through prethermal
  stages~\cite{berges_nonequilibrium_2016,moeckel_interaction_2008,eckstein_thermalization_2009,bertini_prethermalization_2015}
  or be influenced by nonthermal fixed
  points~\cite{berges_nonthermal_2008} or dynamical critical
  points~\cite{heyl_dynamical_2018}, while control of the final
  relaxation process is possible with coherent
  phonons~\cite{zeiger_theory_1992,yang_ultrafast_2014}.
  In this work we study the prethermal state for a single correlated
  band of photoexcited interacting
  electrons~\cite{turkowski_nonlinear_2005,aoki_nonequilibrium_2014}; a multiband case
  was recently discussed in Refs.~\cite{li_nonequilibrium_2021,schuler_gauge_2021}.  For
  a single band we study a Hubbard model with a general time dependence,
  \begin{align}
    \hat{H}(t)
    &=
      \sum_{ij\sigma}
      t_{ij}(t)
      \CD{i\sigma}
      \C{j\sigma}
      +U\sum_i
      \N{i\uparrow}
      \N{i\downarrow}
      \,,\label{eq:hubbardmodel}
  \end{align}
  in terms of the usual fermionic creation, annihilation, and number
  operators for an electron at lattice site $\bm{R}_i$ with spin
  $\sigma$. Here the hopping amplitude $t_{ij}$ is the Fourier
  transform of the dispersion $\epsilon_\BK$, and the Coulomb
  repulsion appears only in the Hubbard interaction $U$.  For weak
  time-dependent interactions $U(t)$ such models exhibit
  prethermalization, i.e., on intermediate time scales a metastable
  state is attained in which quasiparticles are formed, the scattering
  of which then subsequently leads to
  thermalization~\cite{erdos_quantum_2004,schmitt_transient_2008,wais_quantum_2018,mori_thermalization_2018,picano_quantum_2021}. As
  an application, the prethermalization regime can be used to limit
  the heating of periodically driven
  systems~\cite{canovi_stroboscopic_2016,herrmann_floquet_2017,tindall_analytical_2021}.
  Here we study the characteristic features of field-induced
  prethermalized states {for time-dependent but sufficiently weak}
  interactions $U$ $>$ $0$. Our main result is that for a wave train
  containing $m$ pulses with frequency $\omega_\text{pump}$ $=$
  $2\pi/T$ 
  and electric field amplitude $E_{\text{ext}}$, the momentum
  occupation attains a prethermalization plateau after the pulse
  according to
  \begin{align}
    \lim\limits_{m\to\infty}\frac{\langle\hat{n}_{\BK\sigma}\rangle_{t>mT}-\langle\hat{n}_{\BK\sigma}\rangle_{{{0}}}}{mT}
    &=
      e^2 a^2\,s\,
      \frac{\Tilde{n}_{\BK\sigma}^{(2)}(\omega_\text{pump})}{|\omega_\text{pump}+4\pi \IMI
      \sigma(\omega_\text{pump})|^2}
      \,E_{\text{ext}}^2
      \,U^2
      \,,\label{eq:nk-intro-result}
  \end{align}
  for a Hubbard model with diagonal field direction on a hypercubic
  lattice with lattice constant $a$ as defined below
  in~\eqref{eq:H0gauss}; {we set} $\hbar$ $=$ $1$.  This
  scaling limit for long pump pulses involves a function
  $\Tilde{n}_{\BK\sigma}^{(2)}(\omega)$ which is given in
  Section~\ref{sec:secscaledPlateau2} and depends functionally on the
  dispersion, while $s$ is a numerical prefactor of order unity
  depending on the specific envelope, e.g., $s$ $=$
  $1/4$ 
  for our pulse shape~\eqref{eq:field-pulse}.  Pumping with enveloped
  electric field pulses offers more flexibility than interaction
  protocols or continuous driving for the engineering of nontrivial
  metastable states. The denominator in~\eqref{eq:nk-intro-result} is
  due to an additional internal field generated in the sample by the
  external electric field~\cite{skolimowski_misuse_2020} and helps to
  induce the prethermal state when the pump frequency is close to the
  resulting interaction-dependent resonance frequency.

  The paper is organized as follows. In Section \ref{sec:pt} we
  formulate a general weak-coupling approach for electric fields of
  arbitrary strength and use it in Section~\ref{sec:conductivity} to
  obtain the conductivity in lowest order. In
  Section~\ref{sec:prethermal} we discuss the time evolution during
  and after the field pulse in the prethermalization regime. The
  observed scaling with the pulse duration is explained in
  Section~\ref{sec:secscaledPlateau} for the absorbed energy and in
  Section \ref{sec:secscaledPlateau2} for individual momentum
  occupation numbers. In Section~\ref{sec:ppcond} we discuss the
  possibility of distinguishing the prethermal from the thermal state
  by optical spectroscopy and conclude in Section~\ref{sec:conclusion}.
  
  In a gauge with zero electric potential, {the electric field}
  enters only into the hopping amplitudes according to the Peierls
  substitution~\cite{scalapino_insulator_1993}, 
  \begin{align}
    \bm{E}(\bm{r},t)
    &=
      -\frac{1}{c}\frac{\partial\bm{A}(\bm{r},t)}{\partial t}
      \,,~~~~
      t_{ij}(t)
      =
      t_{ij}\,\EXP^{-\tfrac{\IMI e}{{\hbar}
      c}
      \intlimits_{\bm{R}_i}^{\bm{R}_j}\!\D\bm{r} \cdot \bm{A}(\bm{r},t)
      }
      ~\to~
      t_{ij}\,\EXP^{-\tfrac{\IMI e}{{\hbar}
      c}
      ({\bm{R}_i}-{\bm{R}_j})\cdot\bm{A}(t)
      }
      \,,\label{eq:peierls}
  \end{align}
  where in the last step the dipole approximation for the
  long-wavelength limit was used, so that within the sample the field
  $\bm{E}(t)$ $=$ $-\partial_t\bm{A}(t)/c$, $\bm{A}(t)$ $=$
  $A(t)\hat{\bm{a}}$, is approximately homogeneous along a unit vector
  $\hat{\bm{a}}$, and the magnetic field vanishes.  The electric field
  then enters only into the dispersion through $\delta\hat{H}_0(t)$,
  \begin{align}
    \hat{H}_0+\delta\hat{H}_0(t)
    &=
      \sum_{\BK\sigma}
      \epsilon_{\BK-\frac{e}{
      c} \bm{A}(t)}
      \,
      \hat{n}_{\BK\sigma}
      \,.\label{eq:model-Hamiltonian}
  \end{align}
  Apart from the momentum occupation we will study in particular the
  current in the direction of $\hat{\bm{a}}$ and the change of the
  kinetic energy due to the field pulse, starting from the initial
  interacting ground state at time $\tIni$ $=$ $0$,
  \begin{align}
    j(t)
    &=
      \Big\langle{-{\frac{c}{{V}}}}\frac{\partial \hat{H}(t)}{\partial A(t)}
      \Big\rangle_t
      \,,~~~~
      \Delta E_{\text{kin}}(t)
      =
      \langle \hat{H}_0+\delta\hat{H}_0(t) \rangle_{t}-\langle
      \hat{H}_0 \rangle_{{{0}}}
      \,,\label{eq:observables}
  \end{align}
  {where $V$ is the volume.} However, the electric field in the Hamiltonian is not the same as
  the external field impinging on the sample.  According
  to~Ref.~\cite{skolimowski_misuse_2020}, in addition to the latter
  the field which is created in the sample due to Maxwell's equations
  should also be taken into account. The total vector potential
  $\bm{A}(\bm{r},t)$ $=$ $\bm{A}_{\text{ext}}(\bm{r},t)$ $+$
  $\bm{A}_{\text{sys}}(\bm{r},t)$ thus contains an internal part that
  obeys
  $(\partial_t^2-c^2 \nabla^2)\bm{A}_{\text{sys}}(\bm{r},t)= 4\pi c \,
  \bm{j}_{\text{sys}}(\bm{r},t)$. For our case without spatial
  dependence this means $\partial_t^2\bm{A}_{\text{sys}}(t)/(4\pi c)$
  $=$ $\bm{j}_{\text{sys}}(t)$ $=$ $j(t)\hat{\bm{a}}$, where the
  quantum-mechanical expectation value is identified with the
  classical current.  In terms of the linear-response conductivity
  $\sigma(\omega)$ for the internal electric field, the conductivity
  for the external field is then obtained from the partial Fourier
  transform of the current $j(t)$ in the field direction as
  \begin{align}
    j(\omega)
    &=
      E_{\text{ext}}(\omega)\sigma_{\text{ext}}(\omega)
      =
      E(\omega)\sigma(\omega)
      \,,~~~~
      \sigma_{\text{ext}}(\omega)
      =
      \frac{\sigma(\omega)}{\epsilon(\omega)}
      \,,~~~~
      \epsilon(\omega)= 1+\IMI\frac{4\pi}{\omega}\sigma(\omega)
      \,,\label{eq:sigmaExt}
  \end{align}
  {via $j(\omega)$ $=$
    $\IMI\omega E_{\text{sys}}(\omega)/(4\pi)$ and
    $E_{\text{ext}}(\omega)$ $=$ $E(\omega)-E_{\text{sys}}(\omega)$.}
  The total internal field $E(\omega)$ in the
  Hamiltonian~\eqref{eq:peierls} thus equals $E(\omega)$ $=$
  $E_{\text{ext}}(\omega)/\epsilon(\omega)$ {in linear order
    in the field.}
  
  {For our present study, we will start from a given internal
    field pulse and obtain the response of the interacting system to
    all orders in the field, but perturbatively in the interaction.
    For the relation between internal and external field, however,
    only the linear-response connection~\eqref{eq:sigmaExt} will be
    used, for which the required conductivity is obtained
    perturbatively in the interaction in
    subsection~\ref{sec:conductivity}. A more realistic nonlinear
    description of the relation between internal and external field is
    beyond the scope of the present work.}

\begin{figure}[tbp]
    \centering
    \includegraphics[width=0.49\textwidth]{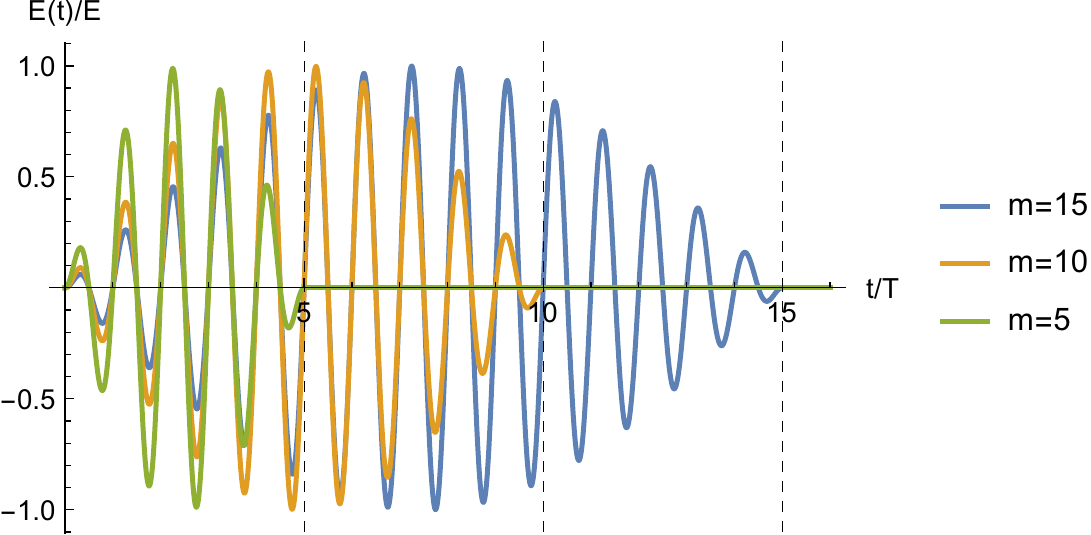}
    \includegraphics[width=0.49\textwidth]{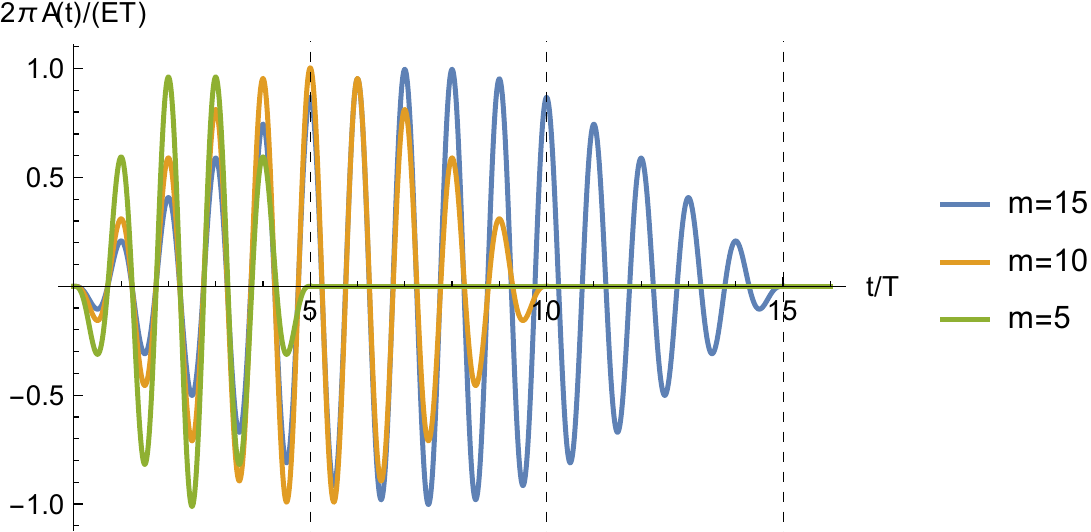}
    \caption{Normalized {dimensionless} electric
      field (left) and vector potential (right).}\label{fig:field}
  \end{figure}

  We assume a pulse for which the Hamiltonian is
  the same before and after the pulse, which is appropriate for a
  metallic system even if the external field were to have different
  vector potentials before and after the pulse.  Specifically, we
  consider a (real-valued) enveloped field pulse for the internal
  electric field with frequency $\omega_\text{pump}$ $=$ $2\pi/T$
  acting from time $\tIni$ $=$ $0$ to $\tFin$ $=$
  $mT$, i.e., a wave train over an integer number $m$ of periods,
  shown in {\textbf{Figure~\ref{fig:field}}},
  \begin{align}
    E(t)
    &=
      -\frac{1}{c}\frac{\partial A(t)}{\partial t}
      =
      E\sin \left( \frac{2\pi t}{T} \right)\sin \left( \frac{\pi
      t}{mT} \right)
      =
      \sum_{j=1}^4
      E_j
      \,
      \EXP^{\IMI t\omega_j}
      \,,\label{eq:field-pulse}
      ~~~~
      E_{\text{ext}}(t)
      =\sum_{j=1}^4
      E_{\text{ext},j}
      \,
      \EXP^{\IMI t\omega_j}
      \,,
  \end{align}
  where $E_{1+2j}$ $=$ $E_{2+2j}$ $=$ $(-1)^{1+j}E/4$ and
  $\omega_{1+2j}$ $=$ $-\omega_{2+2j}$ $=$
  $[1+(-1)^j/(2m)]\,\omega_\text{pump}$
  for $j$ $=$ $0,1$.  
  The  Fourier components $E_j$ of the internal field at
  frequency $\omega_j$ translate into corresponding components
  $E_{\text{ext},j}$ $=$ $\epsilon(\omega_j)E_j$ for the external
  field according to~\eqref{eq:sigmaExt}.

  To illustrate our general results such
  as~\eqref{eq:nk-intro-result}, we will perform explicit evaluations
  following Ref.~\cite{turkowski_nonlinear_2005}, i.e., using
  next-neighbor hopping $t_{ij}$ $=$ $t^*/2\sqrt{d}$ for a hypercubic
  lattice in the limit of infinite dimensions $d$ and assuming a
  diagonal field direction, 
  $\hat{\bm{a}}$ $=$ $(1,1,\ldots,1)$.  {This limit of high
    dimensions~\cite{metzner_correlated_1989} corresponds to dynamical mean-field
    theory, which describes three-dimensional correlated
    electron materials in
    equilibrium~\cite{georges_dynamical_1996,kotliar_electronic_2006}
    and
    nonequilibrium~\cite{aoki_nonequilibrium_2014}
    well in general.  The diagonal field direction is representative for the
    high-dimensional limit, as it does not require further scalings
    with powers of $d$ that would be needed, e.g., for a bond
    direction. The diagonal field direction also leads to technical
    simplifications, as the kinetic energy, the cosine and sine
    dispersions, density of states and joint density of states take
    the following form in the limit of large $d$,}
  \begin{subequations}%
    \label{eq:H0gauss}%
    \begin{align}%
      \hat{H}_0+\delta\hat{H}_0(t)
      &=
        \hat{H}_0 \cos A(t)
        +
        \hat{\mybar{H}}_0 \sin A(t)
        \,,~~~~
        \hat{H}_0
        =
        \sum_{\BK\sigma}
        \epsilon_\BK
        \,
        \hat{n}_{\BK\sigma}
        \,,~~~~
        \hat{\mybar{H}}_0
        =
        \sum_{\BK\sigma}
        \mybar{\epsilon}_\BK
        \,
        \hat{n}_{\BK\sigma}
        \,,\label{eq:H0gauss1}
      \\
      \epsilon_\BK
      &=
        -\frac{t^*}{\sqrt{d}}\sum\limits_{j=1}^{d}\cos (k_ja)    
        \,,~~~~
        \mybar{\epsilon}_\BK
        =
        -\frac{t^*}{\sqrt{d}}\sum\limits_{j=1}^{d}\sin (k_ja)
        \,,\label{eq:H0gauss2}
      \\
      \rho(\epsilon)
      &= \frac1L\sum_\BK \delta(\epsilon-\epsilon_\BK)
        = \frac{\EXP^{-\epsilon^2/{t^*}^2}}{\sqrt{\pi}t^*}
        \,,~~~~
        \rho(\epsilon,\mybar{\epsilon})
        =
        \frac1L\sum_\BK \delta(\epsilon-\epsilon_\BK)\delta(\mybar{\epsilon}-\mybar{\epsilon}_\BK)
        =
        \frac{\EXP^{-(\epsilon^2+\mybar{\epsilon}^2)/{t^*}^2}}{\pi{t^*}^2}
        \,,\label{eq:H0gauss3}
    \end{align}%
  \end{subequations}%
  for $L$ lattice sites. {Here we set $\hbar$, $c$, $a$ to
    unity. We} also set $t^*$ $=$ $1$ and consider only a half-filled
  band, with uncorrelated kinetic energy $E_{\text{kin}}^{(0)}(0)$ $=$
  $-\frac{1}{2\sqrt{\pi}}$ $\simeq$ $-0.282$. {Our units
    are thus $\hbar/t^*$ for time, $t^*/\hbar$ for frequency,
    $\hbar c/(ea)$ for $A(t)$, $t^*/(ea)$ for $E(t)$, $t^*/(a^2\hbar)$
    for $j(t)$, $e^2t^*/(\hbar^2a)$ for $\sigma(t)$, and
    $e^2/(\hbar a)$ for $\sigma(\omega)$.  The term
    $\sigma(\omega)/\omega$ in~\eqref{eq:sigmaExt} requires us to fix
    the scale $\hbar c/(at^*)$; we estimate
    $a\cdot|E_{\text{kin}}^{(0)}(0)|$ $\simeq$ $4$ {\AA} $\cdot$ eV to be
    a representative value and use this in explicit calculations
    involving $\sigma_{\text{ext}}(\omega)$.}  Evaluations for this
  setup can be performed efficiently using integration techniques
  described in the Appendix. In the following we will refer to the
  interacting model~\eqref{eq:hubbardmodel} and~\eqref{eq:H0gauss}
  simply as the Hubbard model with diagonal field direction. We will
  denote $\hat{n}_{\BK\sigma}$ as
  $\hat{n}_{\epsilon\mybar{\epsilon}\sigma}$ when it depends only on
  $\epsilon_\BK$ and $\mybar{\epsilon}_\BK$.

  \subsection{Weak-coupling theory}\label{sec:pt}

  For the initial and time-evolved states of an interacting
  Hamiltonian we use a perturbative formulation which is also useful
  for prethermalization phenomena after general interaction protocols
  $U(t)$ as discussed separately
  elsewhere~\cite{alexander_notitle_nodate}. For a general
  time-dependent Hamiltonian $\hat{H}(t)$ $=$ $\hat{H}_0(t)$ $+$
  $g\hat{V}(t)$, an operator $\hat{A}$ evolved in the interaction
  picture reads
  \begin{align}
    \hat{S}^{\dagger}(t)\hat{A}_I({ t})\hat{S}(t)
    =
    \sum\limits_{n=0}^{\infty} (ig)^n
    \!\intlimits_{0}^{t}\!\D t_1
    \ldots
    \!\intlimits_{0}^{t_{n-1}}\!\D t_n
    \big[ \hat{V}_{I}(t_n), \ldots
    \big[ \hat{V}_{I}(t_1) , \hat{A}_I({t})
    \big]
    \ldots \big]
    \label{exactDirac}\,,
  \end{align}
  with the interaction picture propagator $\hat{S}(t)$ $=$
  $\hat{U}^{\dagger}_0(t)\hat{U}(t)$, defined in terms of the
  propagators $\hat{U}_0(t)$ and $\hat{U}(t)$ of $\hat{H}_0(t)$ and
  $\hat{H}_0(t)$ $+$ $g\hat{V}(t)$, respectively, and $\hat{O}_I(t)$
  $=$ $\hat{U}^{\dagger}_0 (t) \hat{O}(t)\hat{U}_0(t)$ for any
  Schr\"odinger operator $\hat{O}(t)$.

  For our Hamiltonian~\eqref{eq:model-Hamiltonian}, the initial
  interacting ground state of~$\hat{H}(0)$ $=$ $\hat{H}_0$ $+$
  $g\hat{V}$ (where $g$ $=$ $U$ and $\hat{V}$ $=$
  $\sum_i \N{i\uparrow} \N{i\downarrow}$) at $t$ $=$ $0$ is
  time-evolved with the additional kinetic energy
  term~$\delta\hat{H}_0(t)$, which vanishes before and after the
  electric field pulse.  We consider an auxiliary Hamiltonian in which
  the interaction term is switched on adiabatically for negative
  times, so as to generate the interacting initial state from the
  corresponding noninteracting eigenstate $|\Psi(t=-\infty)\rangle$
  $=$ $|\Psi_0\rangle$ with $\hat{H}_0|\Psi_0\rangle$ $=$
  $E_0 |\Psi_0\rangle$ and expectation values
  $\langle\cdots\rangle_{(0)}$; by contrast expectation
  values in the initial interacting eigenstate are denoted by
  $\langle\cdots\rangle_{t=0}$ $=$
  $\langle\cdots\rangle_{{{0}}}$. Namely we set
  \begin{align}
    \hat{H}(t)=\hat{H}_0+\delta\hat{H}_0(t) + g\,f(t)\,\hat{V}
    \,,~~~~
    f(t)=
    \begin{cases}
      \EXP^{\delta t} & \text{if~} t<0,
      \\
      1 & \text{if~}t\geq0,
    \end{cases}
          ~~~~
          \delta\to 0^+
          \,.
  \end{align}
  Here $\hat{H}_0+\delta \hat{H}_0(t)$ has the same set of eigenstates
  $\{\ket{\Psi_l}\}$ as $\hat{H}_0$. The noninteracting propagator for
  $t$ $\geq$ $0$ is therefore simply $\hat{U}_0(t)$ $=$
  $\exp(-\IMI\int_{0}^{t}\D\tau\, [\hat{H}_0+\delta\hat{H}_0(\tau)])$.
  Expanding in the interaction strength $g$ we have for any observable
  $\hat{A}$,
  \begin{subequations}%
    \label{eq:PulseInteraction}%
    \begin{align}%
      \langle \hat{A} \rangle_t
      &=
        \langle \hat{A} \rangle_{(0)} +g\Delta A^{(1)}(t) +O(g^2)
        \,,\\
      \Delta A^{(1)}(t)
      &=
        \IMI\intlimits_{-\infty}^{t}\!\D t_1\,f(t_1) \langle \big[ \hat{V}_{I}(t_1) , \hat{A}_I({t}) \big] \rangle_{(0)}
        =
        \sum_l V_{0l}\,A_{l0}\,\varphi_l^{(1)}(t)+\text{c.c.}
        \,,\\
      \varphi_l^{(1)}(t)
      &=
        \IMI\intlimits_{-\infty}^{t}\!\D t_1 \, f(t_1)\EXP^{\IMI\intlimits_{t_1}^{t}\!\D\tau\,\Delta E_l+\delta e_l(\tau)}
        \label{eq:extractPhi1}
        =
            -\frac{1}{\Delta E_l} + \IMI\intlimits_{0}^{t}\!\D t_1\, \EXP^{\,i(t-t_1) \Delta E_l} \big(\EXP^{\IMI\int_{t_1}^{t}\!\D\tau\, \delta e_l(\tau)}-1\big)
            \,,
    \end{align}%
  \end{subequations}%
  where we used the abbreviations $\bra{\Psi_l}\hat{A}\ket{\Psi_m}$
  $=$ $A_{lm}$, $\Delta E_l$ $=$ $E_l-E_0$,
  $\delta\hat{H}_0(t)\ket{\Psi_l}$ $=$ $\delta E_l(t)\ket{\Psi_l}$,
  $\delta e_l(t)$ $=$ $\delta E_l(t)-\delta E_0(t)$.  We label
  observables that commute with $\hat{H}_0$ as second-order
  observables $\hat{a}$ (because the first-order term vanishes for
  them), otherwise as first-order observables $\hat{A}$. For a
  second-order observable we have
  \begin{subequations}%
    \begin{align}%
      \langle \hat{a} \rangle_t
      &=
        \langle \hat{a} \rangle_{(0)} +g^2\Delta a^{(2)}(t)+O(g^3)
        \,,\\
      \Delta a^{(2)}(t)
      &=
        -\intlimits_{-\infty}^{t}\!\D t_1 \intlimits_{-\infty}^{t_{1}}\!\D t_2\,f(t_1)f(t_2)\langle\big[ \hat{V}_{I}(t_2), \big[ \hat{V}_{I}(t_1) , \hat{a} \big] \big]\rangle_0
        = \sum_l |V_{0l}|^2 \Delta a_l \varphi_l^{(2)}(t)
        \,,\label{eq:extractPhi2}
      \\
      \varphi_l^{(2)}(t)
      &=
        \intlimits_{-\infty}^{t}\!\D t_1 \intlimits_{-\infty}^{t_{1}}\!\D t_2\, f(t_1)f(t_2)\EXP^{\IMI\intlimits_{t_2}^{t_1}\!\D\tau\,\Delta E_l+\delta e_l(\tau)}+\text{c.c.}
        \nonumber\\
      &=\frac{1}{\Delta E_l^2} -2\,\RE\intlimits_{0}^{t}\!\D t_1 \intlimits_{0}^{t_1}\!\D t_2\, \exp\Big( \IMI\intlimits_{t_2}^{t_1}\!\D t'\,(\Delta E_{l}+\delta e_{l}(t')) \Big) \frac{\delta e_{l}(t_2)}{\Delta E_{l}}
        \,,
        \label{eq:PulsOthers}
    \end{align}%
    \label{eq:secondorderobservables}%
  \end{subequations}%
  where $\hat{a}\ket{\Psi_l}=a_l\ket{\Psi_l}$, and
  $\Delta a_l=a_l-a_0$. Without electric field, $\delta e_l(t)=0$,
  both~\eqref{eq:PulseInteraction} and~\eqref{eq:secondorderobservables} reduce to
  the standard perturbative result for the interacting ground state.

  \subsection{{Relation between internal and external field in linear response}}
  \label{sec:conductivity}

  {As discussed in the first subsection, below we will use a
    given internal field pulse~\eqref{eq:field-pulse}, which is
    related to the external field pulse according
    to~\eqref{eq:sigmaExt} in linear order in the field. We therefore
    use the expressions of the previous subsection to obtain the
    conductivity in second order in the interaction.  Expanding} to
  $O(A^2)$ with $\bm{A}(t)$ $=$ $A(t)\hat{\bm{a}}$ gives
  {(with $\hbar$, $c$, $a$ set to unity)}
  \begin{align}
    \!\!\!\!\!\!\!\!
    \hat{j}(t)
    =
      -\frac{1}{{L}}\frac{\partial \hat{H}(t)}{\partial A(t)}
      =
    e
      \hat{H}_0^{(1)} - {e^2}
    A(t) \hat{H}_0^{(2)}+O(A^2)
      \,,\;
      \hat{H}_0^{(n)}
      =
      {\frac{1}{L}}
      \sum_{k\sigma}\epsilon^{(n)}_{\BK}\hat{n}_{\BK\sigma}
      \,,\;
      \epsilon^{(n)}_{\BK}=\frac{\partial^n\epsilon_{\BK+\hat{\bm{a}}{x}}}{\partial x^n }\Big|_{x=0}
      \,.\!\!\!
  \end{align}
  For a field that is zero before $\tIni$ $=$ $0$, the
  linear-response result for the current, $j(t)$ $=$
  $\langle \hat{j}\rangle_t$, then comprises the usual diamagnetic and
  paramagnetic contribution to the conductivity,
  \begin{subequations}%
    \begin{align}%
      j(t)
      &=
        \intlimits_0^{t}\!\D\tau \,E(\tau)\sigma(t,\tau)+O(E^2)
        \,,~~
        \sigma(t,\tau)= \sigma^{\text{dia}}(t)+ \sigma^{\text{pm}}(t,\tau)\,,
        \label{eq:CurrentResponse-general}
      \\
      \sigma^{\text{dia}}(t)
      &=
        {e^2}
        \,
        \langle\hat{H}_0^{(2)}\rangle_{t}
        \,,~~    
        \sigma^{\text{pm}}(t,\tau)
        =
        \IMI
        {e^2}
        \intlimits_0^{t}\!\D\tau'\,
        \langle[\hat{H}_0^{(1)}(\tau'),\hat{H}_0^{(1)}(t)]
        \rangle_{{{0}}}\,.
    \end{align}\label{eq:currentResponse}%
  \end{subequations}%
  where the time dependences of $\hat{H}_0^{(1)}$ are in the
  Heisenberg picture of the Hamiltonian without field.  If that
  Hamiltonian is time-independent this simplifies to
  $\sigma^{\text{dia}}(t)$ $\to$ $\sigma^{\text{dia}}$,
  $\sigma^{\text{pm}}(t,\tau)$ $\to$ $\sigma^{\text{pm}}(t-\tau)$, and
  $j(\omega)$ $=$ $E(\omega)\sigma(\omega)$.
  For the interacting Hamiltonian $\hat{H}(0)$ $=$ $\hat{H}_0$ $+$
  $g\hat{V}$ we use~\eqref{eq:secondorderobservables} to find in the leading orders in~$g$ that
  \begin{subequations}%
    \begin{align}%
      \sigma^{\text{dia}}
      &=
        \sigma^{\text{dia},(0)}+g^2\sigma^{\text{dia},(2)}+O(g^3)
        \,,~~
        \sigma^{\text{dia},(0)}
        =
        {e^2}
        \,
        \langle\hat{H}_0^{(2)}\rangle_{0}
        \,,~~
      \sigma^{\text{dia},(2)}
      =
      {e^2}
      \sum_l |V_{0l}|^2 \frac{\Delta E_l^{(2)} }{\Delta E_l^2}
      \,,\!\!
      \\
      \!\!
      \!\!
      \!\!
      \sigma^{\text{pm}}(t-\tau)&=g^2\sigma^{\text{pm},(2)}(t-\tau)+O(g^3)
                                  \,,~
                                  \sigma^{\text{pm},(2)}(t)
                                  =
                                  {e^2}
                                  \sum_l |V_{0l}|^2 2(\Delta E^{(1)}_l)^2\,\frac{\cos (\Delta E_l t)-1}{\Delta E_l^3}\,.
    \end{align}%
  \end{subequations}%
  with
  $\hat{H}_0^{(n)}\ket{\Psi_l}=E_l^{(n)}\ket{\Psi_l}$ and
  $E_l^{(n)}-E_0^{(n)}=\Delta E_l^{(n)}$, and the constant $\sigma^{\text{dia}}$ leading to the familiar
  Drude peak in the partial Fourier transform,
  $\sigma^{\text{dia}}(\omega)$ $=$
  $\IMI\sigma^{\text{dia}}/(\omega+\IMI\delta)$, $\delta$ $\to$ $0^+$.

\begin{figure}[tbp]
    \centering
    \includegraphics[width=0.49\textwidth]{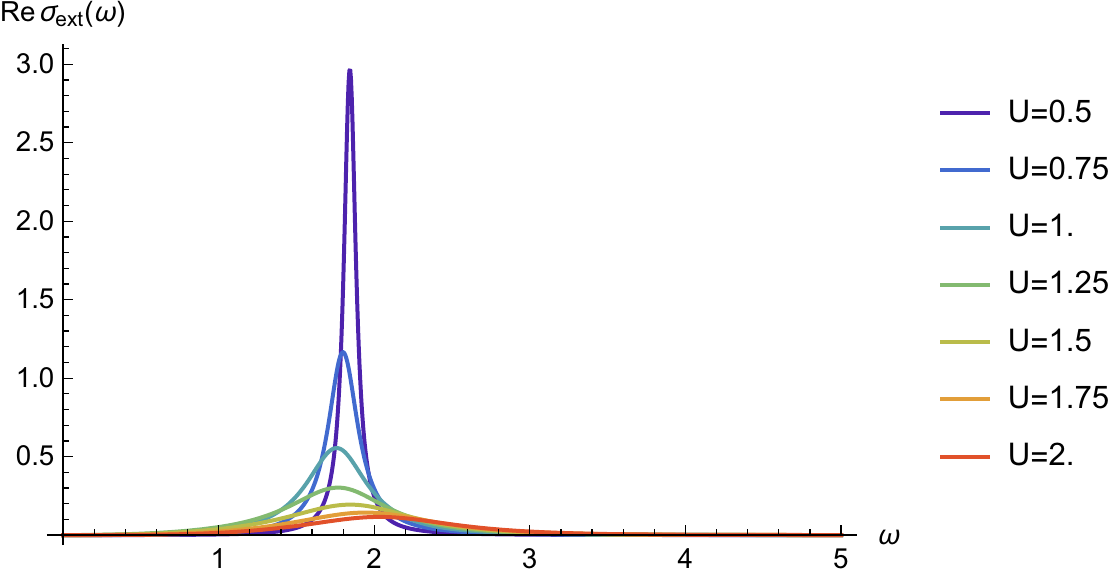}
    \includegraphics[width=0.49\textwidth]{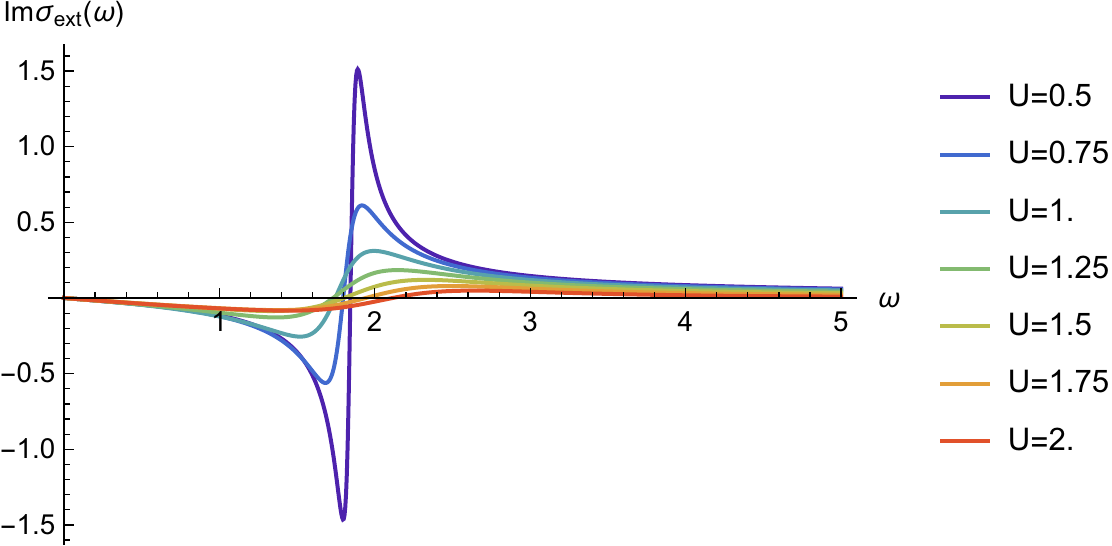}
    \caption{Optical conductivity $\sigma_{\text{ext}}$ {with respect
        to the external field according to}~\eqref{eq:sigmaExt} for
      the Hubbard model with diagonal field
      direction~\eqref{eq:H0gauss}, as given
      in~\eqref{eq:conductivityHubbard}.
      Here $\sigma_{\text{ext}}$ is in units of
        $e^2/(\hbar a)$ and $\omega$ in units of $t^*/\hbar$, and we have
        chosen $a\cdot|E_{\text{kin}}^{(0)}(0)|$ $=$ $4$ {\AA} $\cdot$ eV, see
        remarks below~\eqref{eq:H0gauss}.
      }
    \label{fig:Conductivity}
  \end{figure}

  We evaluate these expressions for the Hubbard model with diagonal
  field direction~\eqref{eq:H0gauss}, using weight functions~$a_n(b)$
  to represent Gaussian integrals as described in~\eqref{eq:an} of
  the Appendix. For coupling $g$ $=$ $U$ and with
  ${e^2}$ $=$ $1$, $t^*$ $=$ $1$, we have
  $\hat{H}_0^{(2)}=-\hat{H}_0$ and
  $\hat{H}_0^{(1)}=\hat{\mybar{H}}_0$, leading to
  \begin{subequations}%
    \label{eq:conductivityHubbard}%
    \begin{align}%
      \sigma(\omega)
      &=
        \IMI\frac{\sigma^{\text{dia},(0)}+U^2\sigma^{\text{dia},(2)}}{\omega+\IMI\delta}
        +U^2\sigma^{\text{pm}(2)}(\omega)+O(U^3)
        \,,
      \\
      \sigma^{\text{dia},(0)}
      &=
        \frac{1}{2\sqrt{\pi}}
        \simeq
        0.282
        \,,~~~~
        \sigma^{\text{dia},(2)}
        =-\intlimits_{\frac{1}{4}}^{1}\!\D b\, \frac{a_4(b)}{2\sqrt{\pi^3 b^3}}\simeq -0.0659
        \,,\label{eq:conductivityDIA}
      \\
      \sigma^{\text{pm},(2)}(\omega)
      &=
        \intlimits_{\frac{1}{4}}^{1}\!\D b\, \frac{2a_4(b)}{\sqrt{b \pi^3}}
        \bigg(\frac{-\IMI}{\omega+\IMI\delta}+\sqrt{\pi b}\,\EXP^{-\omega^2 b}+2\IMI\sqrt{b}\,\DawsonF(\sqrt{b}\omega)\bigg)
        \,,\label{eq:conductivityPM}
    \end{align}%
  \end{subequations}%
  in terms of the Dawson function $\DawsonF(x)$ $=$
  $\EXP^{-x^2}\int_{0}^{x}\!\D t\,\EXP^{t^2}$ and the weight function
  $a_4(b)$ of~\eqref{eq:a4}. The corresponding optical conductivity {$\sigma_{\text{ext}}(\omega)$}
  for the external field is then calculated by inserting this result
  into~\eqref{eq:sigmaExt} and is shown in
  {\textbf{Figure~\ref{fig:Conductivity}} with parameters as given below~\eqref{eq:H0gauss}}.
  In its denominator we keep $\epsilon(\omega)$ as a series in $U$ and
  do not expand it into the numerator. Similar to the result for the
  absorbed power~\cite{skolimowski_misuse_2020}, this denominator
  suppresses the zero-frequency pole for finite $U$ and turns it into
  a resonance at $\omega_*$, which approaches $\omega_*$ $=$
  $\sqrt{2\sqrt{\pi}}$ $\simeq$ $1.88$ (in units of $t^*$ $=$ $1$) in
  the noninteracting limit.

  \section{Nonperturbative effects of a pump pulse with finite duration}

  \subsection{Pulse-induced transient states and prethermalization}
  \label{sec:prethermal}

\begin{figure}[b]
    \centering
    \includegraphics[width=0.49\textwidth]{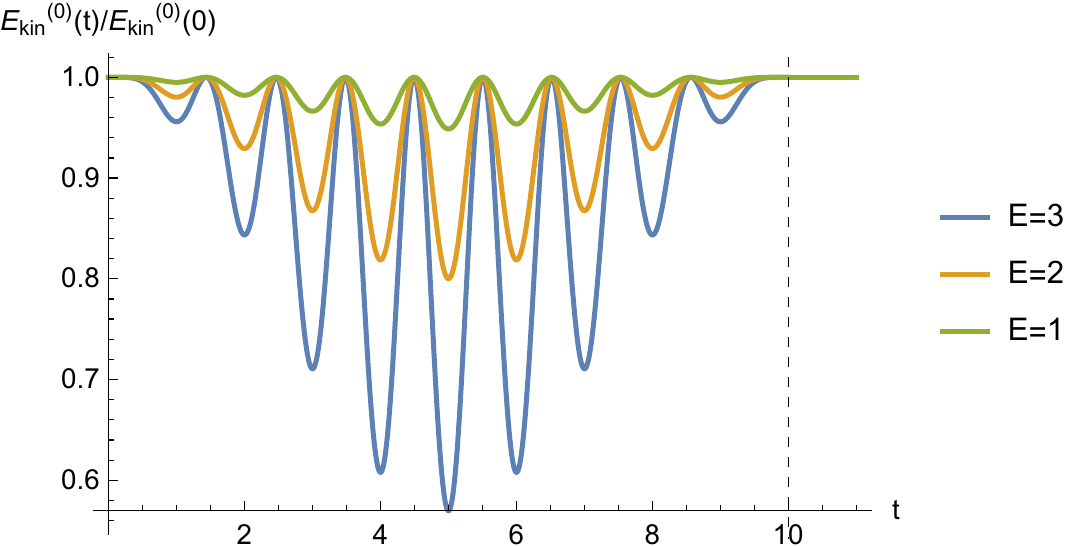}
    \includegraphics[width=0.49\textwidth]{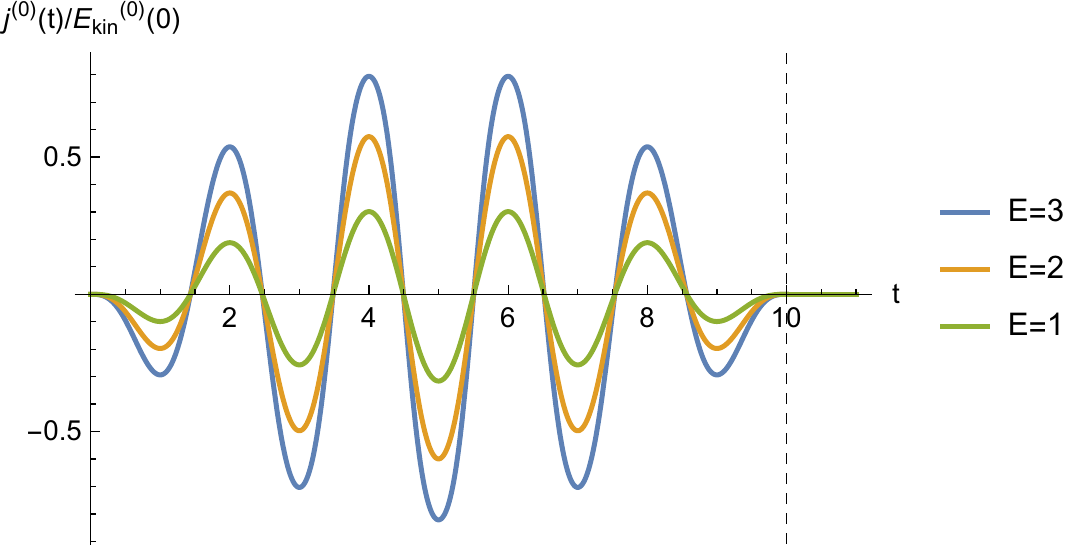}
    \caption{Normalized kinetic energy (left) and current (right) for
      the noninteracting ($U$ $=$ $0$) Hubbard model with diagonal field
      direction~\eqref{eq:H0gauss} subjected to the pump
      pulse~\eqref{eq:field-pulse} with $T$ $=$ $2$, $m$ $=$ $5$.
      Here the units are $\hbar/t^*$ for time,
        $t^*/(a^2\hbar)$ for current, $t_*$ for energy, and $t^*/(ea)$
        for electric field, see
        remarks below~\eqref{eq:H0gauss}.
    }\label{fig:Obs0}
  \end{figure}

\begin{figure}[t]
    \centering
    \includegraphics[width=0.49\textwidth]{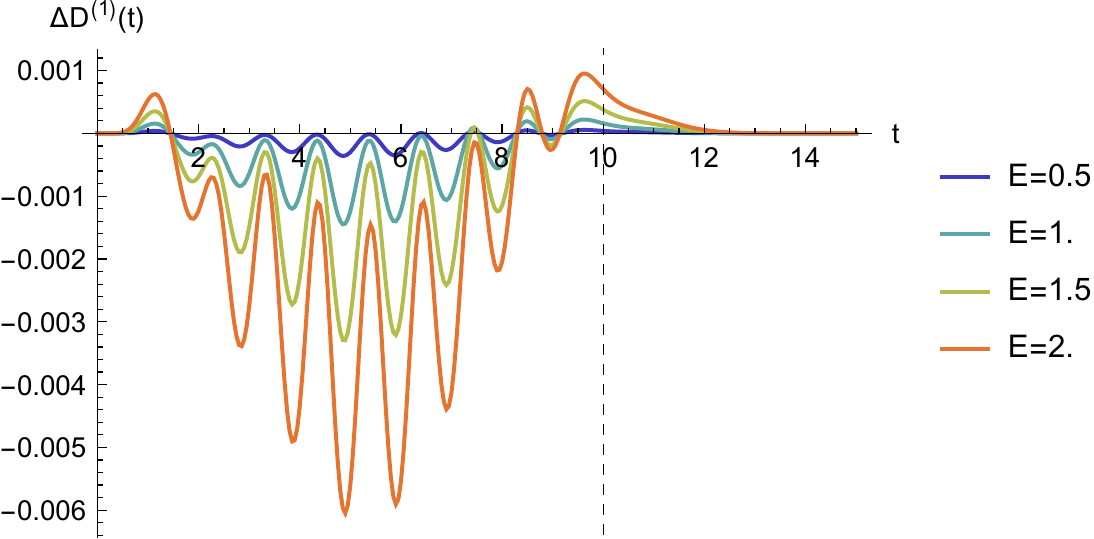}
    \includegraphics[width=0.49\textwidth]{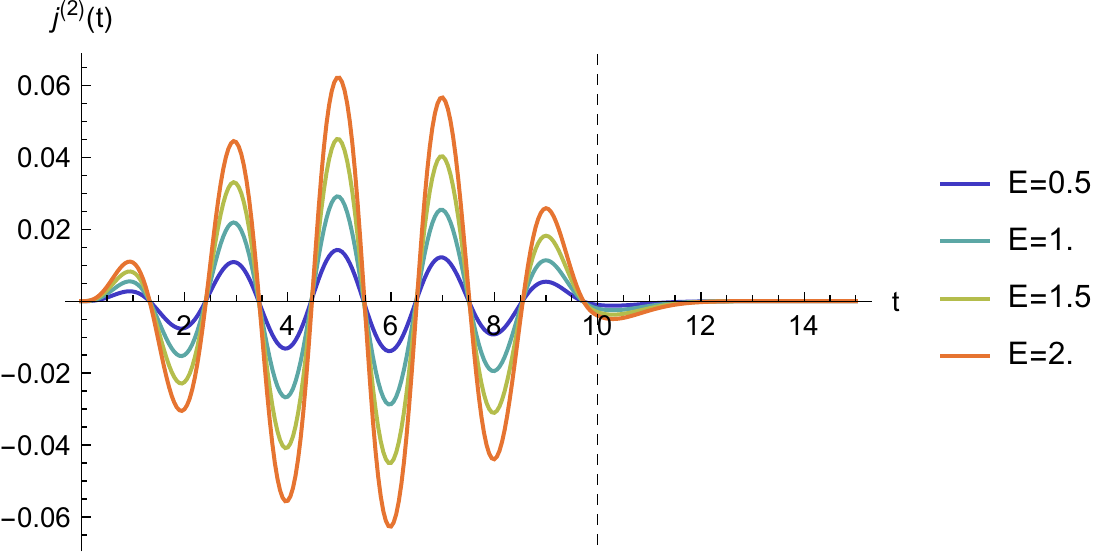}
    \includegraphics[width=0.49\textwidth]{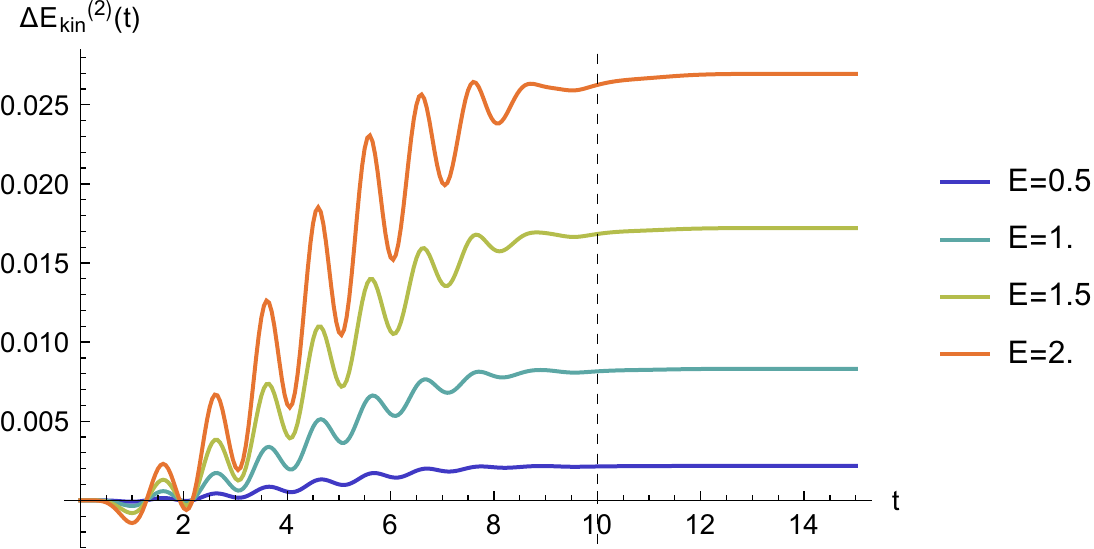}
    \includegraphics[width=0.49\textwidth]{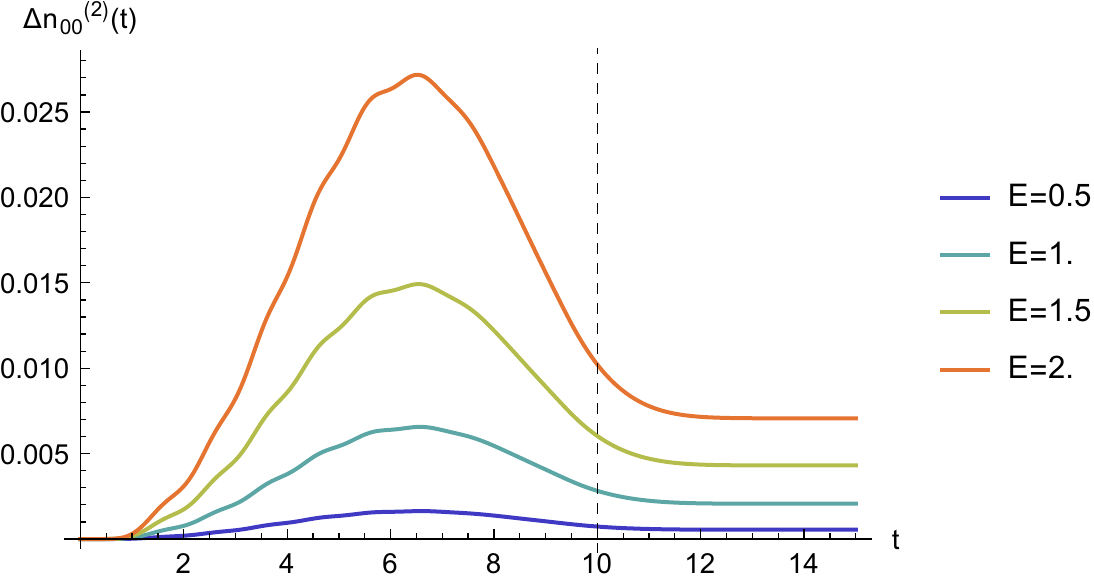}
    \caption{Leading contributions
      {(in order $U^n$ as indicated by the upper index $^{(n)}$)}
      to the change in double occupation
      (top left), current (top right), kinetic energy (bottom left)
      and momentum occupation number with
      $\epsilon_\BK$ $=$ $\mybar{\epsilon}_\BK$ $=$ $0$ (bottom right) for the
      Hubbard model with diagonal field direction~\eqref{eq:H0gauss}
      subjected to the pump pulse~\eqref{eq:field-pulse} with $T$ $=$ $2$,
      $m$ $=$ $5$.
      Here the units are $\hbar/t^*$ for time,
        $t^*/(a^2\hbar)$ for current, $t_*$ for energy, and $t^*/(ea)$
        for electric field, see
        remarks below~\eqref{eq:H0gauss}.
    }\label{fig:ObsVara}
  \end{figure}

  We consider the enveloped pump pulse $E(t)$
  of~\eqref{eq:field-pulse}, which is shown in Figure~\ref{fig:field}
  together with its vector potential $A(t)$. For simplicity we will
  focus on the effect of this given \emph{internal} field $E(t)$ on an
  interacting system, as the relation to the external field
  $E_{\text{ext}}(t)$ also involves the interaction according
  to~\eqref{eq:sigmaExt}-\eqref{eq:field-pulse}.  Because the momentum
  occupation numbers $\langle \hat{n}_{\BK\sigma}\rangle_t$ are
  second-order observables we expand the observables $j(t)$ and
  $E_{\text{kin}}(t)$ of~\eqref{eq:observables} by means
  of~\eqref{eq:secondorderobservables}, $E_{\text{kin}}(t)$ $=$
  $E_{\text{kin}}^{(0)}+g^2E_{\text{kin}}^{(2)}(t)+O(g^3)$, $j(t)$ $=$
  $j^{(0)}(t)+g^2j^{(2)}(t)+O(g^3)$.  For the Hubbard model with
  diagonal field direction~\eqref{eq:H0gauss} the time-dependent
  zeroth-order terms of kinetic energy and current are depicted in
  \textbf{Figure~\ref{fig:Obs0}}.
  After the pulse they have returned
  to their initial values as the momentum occupation numbers remain
  constant in the noninteracting case. In the presence of
  interactions, the electric field induces changes that are depicted
  in \textbf{Figure~\ref{fig:ObsVara}},
  in which the change in the
  double occupation, $\Delta D(t)$ $=$
  $\langle\N{i\uparrow}\N{i\downarrow}\rangle_t$ $-$
  $\langle\N{i\uparrow}\N{i\downarrow}\rangle_0$, a first-order
  observable, is also plotted. The averaged quantities follow the
  electric field amplitude closely. {Individual momentum 
    occupation numbers are not gauge independent during the pulse~\cite{aoki_nonequilibrium_2014}; for
    our gauge they} show slowly varying behavior with slight modulations
  during one period $T$.  We further note that even for the very
  strong fields $E\geq1$ considered here, the approximate field
  dependence is linear in $E$ for $j^{(2)}(t)$ and quadratic for
  $D^{(1)}(t)$ and $E_{\text{kin}}^{(2)}(t)$, which will be further
  studied in the next subsection.  At the end of the field pulse these
  quantities undergo a further relaxation. As a first-order observable
  the double occupation relaxes to its value prior to the pulse, while
  the current relaxes to zero as it is a first-order observable in the
  electric field. The kinetic energy and the momentum occupation
  numbers are second-order observables in interaction strength and in
  the electric field and as such relax to a finite value on a
  time-scale on the order of $1/t_*$ $\equiv$ $1$. At second order in
  the interaction this is the prethermalization regime during which
  quasiparticles are formed, analogous to the case of time-dependent
  interaction
  protocols~\cite{berges_nonequilibrium_2016,erdos_quantum_2004,moeckel_interaction_2008},
  and the kinetic energy and double occupation are already thermalized
  on this time scale. Further relaxation of individual momentum
  occupation numbers is expected due to the scattering of
  quasiparticles, but which we do not consider here.

  \subsection{Scaling behavior of the absorbed energy}
  \label{sec:secscaledPlateau}

\begin{figure}[t]
    \centering
    \includegraphics[width=0.49\textwidth]{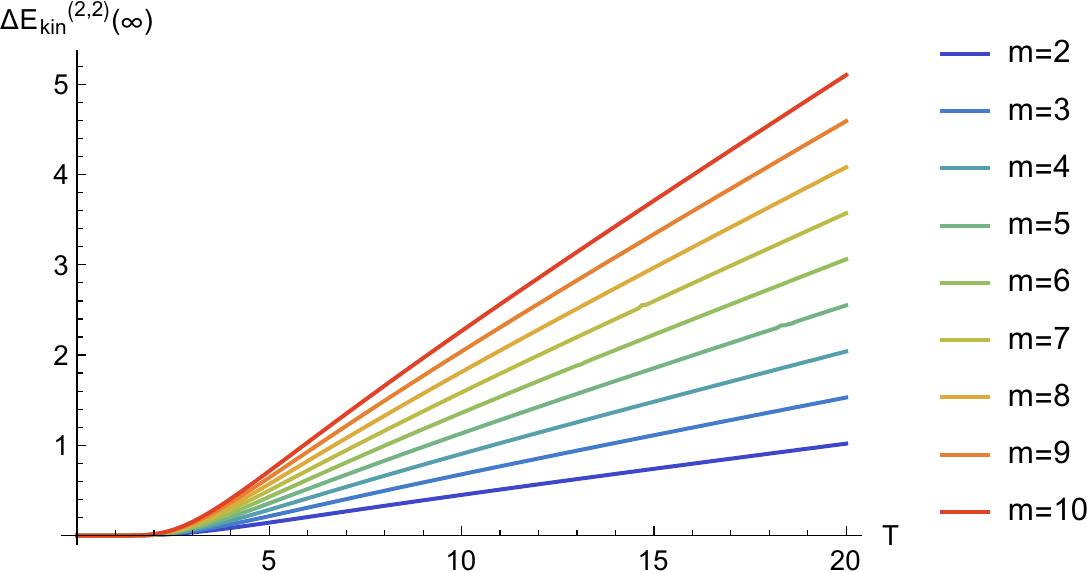}
    \includegraphics[width=0.49\textwidth]{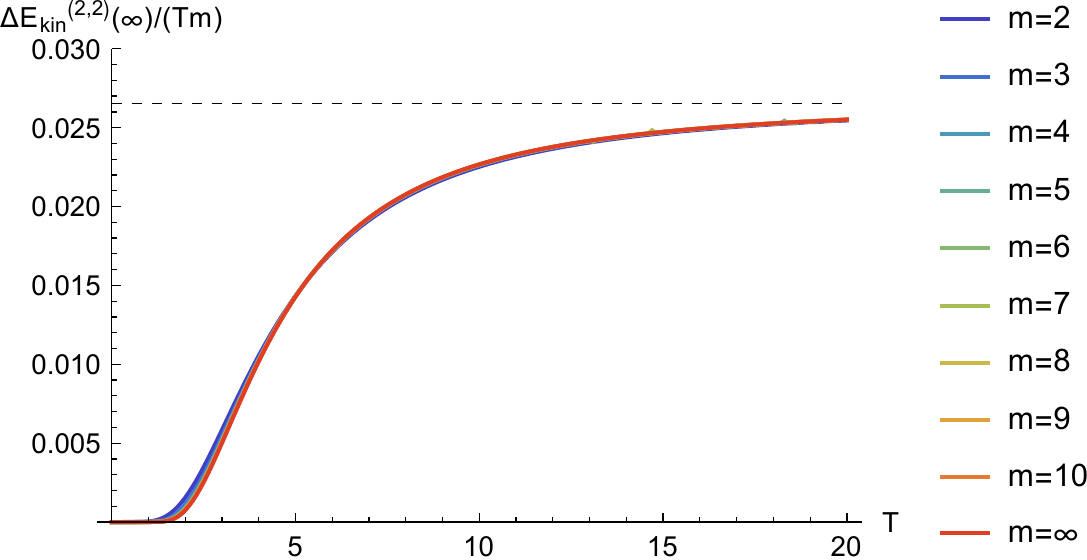}
    \caption{Prethermalization plateau of the kinetic energy {(in order $U^2E^2$)} for the
      Hubbard model with diagonal field direction~\eqref{eq:H0gauss}
      subjected to the pump pulse~\eqref{eq:field-pulse}, unscaled
      (left) and scaled by pulse duration (right) for varying number
      of pulse oscillations $m$ and periods
      $T$. The limit of long pulse durations ($m$ $=$ $\infty$)
      corresponds to~\eqref{eq:heatingSpecific}.
      Here the units are $\hbar/t^*$ for time,
        $t_*$ for energy, and $t^*/(ea)$
        for electric field, see
        remarks below~\eqref{eq:H0gauss}.
    }\label{fig:H0plateauScale}
  \end{figure}

  From Figures \ref{fig:Obs0} and~\ref{fig:ObsVara} it is apparent
  that the electronic response scales approximately with the field
  amplitude, as we now analyze in further detail. Since the double
  occupation eventually returns to its initial value, the change in
  kinetic energy corresponds to the absorbed electric field energy.
  Its leading term in the electric field,
  $\Delta E_{\text{kin}}^{(2)}(t)$ $=$
  $E^2 \Delta E_{\text{kin}}^{(2,2)}(t)+O(E^3)$ has a long-time limit
  that exhibits an approximate linear scaling with the pulse duration
  $\tFin$ $=$ $mT$ as shown in
  \textbf{Figure~\ref{fig:H0plateauScale}}.
  All the plotted
  prethermalization plateaus
  $\Delta E_{\text{kin}}^{(2,2)}(\infty)/mT$ collapse quite well onto
  a single curve, which should thus be described by the absorbed
  energy in the limit $m\to\infty$ of long pulse durations, which we
  now calculate. In leading order in the field we obtain for the
  absorbed field energy,
  \begin{subequations}%
    \label{eq:absorptionExpansion}
    \begin{align}%
      \delta\hat{H}_0(t)
      &=-\frac{{e}}{c}
        A(t)\hat{H}_0^{(1)}+O(A^2)\,,\\
      \Delta\langle\hat{H}\rangle_t
      &=
        \frac{-1}{c^2}
        \intlimits_{0}^{t}\!\D t_1
        \intlimits_{0}^{t_1}\!\D t_2
        \,A(t_1)A(t_2)
        \,
        \langle
        {e^2}
        \,
        \big[\hat{H}_0^{(1)}(t_2),\big[\hat{H}_0^{(1)}(t_1),\hat{H}\big]\big]
        \rangle_{{{0}}}
        +O(A^3)
        \,,
    \end{align}%
  \end{subequations}%
  for a general interaction, 
  in which we recognize the expectation value as
  $\partial^2\sigma^{\text{pm}}(t)/\partial t^2|_{t=t_1-t_2}$,
  involving the paramagnetic conductivity
  of~\eqref{eq:currentResponse}; as in that equation, the time
  dependences of $\hat{H}_0^{(1)}$ in~\eqref{eq:absorptionExpansion}
  are in the Heisenberg picture of the Hamiltonian without field.  
  To establish the approximate scaling with the pulse duration, we
  use a Fourier representation and consider only times $t$ after the
  end of the pulse,
  \begin{align}
    \Delta\langle\hat{H}\rangle_{t>\tFin}
    &=
      \int\!\D\omega\,
      \frac{\omega^2\,\RE\{\sigma^{\text{pm}}(\omega)\}}{c^2\pi}
      \intlimits_{0}^{\tFin}\!\D t_1
      \intlimits_{0}^{t_1}\!\D t_2
      \,A(t_1)A(t_2)\EXP^{\IMI (t_1-t_2)(\omega+\IMI\delta)} +O(A^3)\,,
  \end{align}
  in which real part of
  the partial Fourier transform $\sigma^{\text{pm}}(\omega)$ appears
  since $\sigma^{\text{pm}}(t)$ is real and symmetric. For any
  (real-valued) electric field pulse that can be decomposed as
  $E(t)$ $=$ $\sum_j E_j \EXP^{\IMI t\omega_j}$ as
  in~\eqref{eq:field-pulse}, we obtain in leading order 
  in the pulse duration $\tFin$ that
  \begin{align}
    \frac{1}{c^2}
    \intlimits_{0}^{\tFin}\!\D t_1
    \intlimits_{0}^{t_1}\!\D t_2\,
    A(t_1)A(t_2)\EXP^{\IMI
    (t_1-t_2)(\omega+\IMI\delta)}=\sum_j
    \frac{|E_j|^2}{\omega_j^2}\frac{\IMI \tFin}{\omega+\IMI\delta-\omega_j}+O(\tFin^0)
  \end{align}
  Performing the limit $\delta$ $\to$ $0^+$ and collecting the delta
  contributions $(\omega+\IMI\delta-\omega_j)^{-1}\to -\IMI\pi
  \delta(\omega-\omega_j)$ gives us the following general result for
  the scaled long-time limit of the absorbed energy,
  \begin{align}
    \lim\limits_{\tFin\to\infty}
    \frac{\Delta\langle\hat{H}\rangle_{t>\tFin}}{\tFin}
    &=
      \sum_j
      |E_j|^2\,\RE\{\sigma^{\text{pm}}(\omega_j)\}
      +O(E^3)\,.\label{eq:heatingGeneral}
  \end{align}
  This result for the absorbed power is still independent of the interaction.
  For the wave train~\eqref{eq:field-pulse} with $m$ pulses
  of period $T$ $=$ $2\pi/\omega_\text{pump}$ and in leading order in the interaction it becomes
  \begin{align}
    \lim\limits_{m\to\infty}
    \frac{\Delta\langle\hat{H}\rangle_{t>mT}}{mT}
    =
    \frac{1}{4}\,E^2\,U^2\,\RE\{\sigma^{\text{pm},(2)}(\omega_\text{pump})\}
    +O(E^3U^2)+O(E^2U^3)\,,\label{eq:heatingSpecific}
  \end{align}
  where the numerical prefactor is particular to our specific pulse
  shape.  In Figure~\ref{fig:H0plateauScale} this expression is
  plotted with the label $m$ $=$ $\infty$ and calculated from the
  paramagnetic conductivity in~\eqref{eq:conductivityPM} for the
  Hubbard model with diagonal field direction~\eqref{eq:H0gauss}.  We
  conclude that for sufficiently long pulses, the absorbed power is
  well approximated by the leading orders in field and interaction for
  all periods $T$.
  Relations similar
  to~\eqref{eq:heatingGeneral}-\eqref{eq:heatingSpecific} between
  the absorbed power and the conductivity for essentially continuous
  pulses were also discussed
  in~\cite{skolimowski_misuse_2020}. In the limit of long pulse
  durations we can  replace the internal field amplitude $E$ in~\eqref{eq:heatingSpecific}  by
  $E_{\text{ext}}(\omega_\text{pump})/\epsilon(\omega_\text{pump})$
  to obtain the dependence on the external field, as further
  discussed in the next subsection.

  \subsection{Scaling behavior of the momentum distribution}
  \label{sec:secscaledPlateau2}

\begin{figure}[t]
    \centering
    \includegraphics[width=0.49\textwidth]{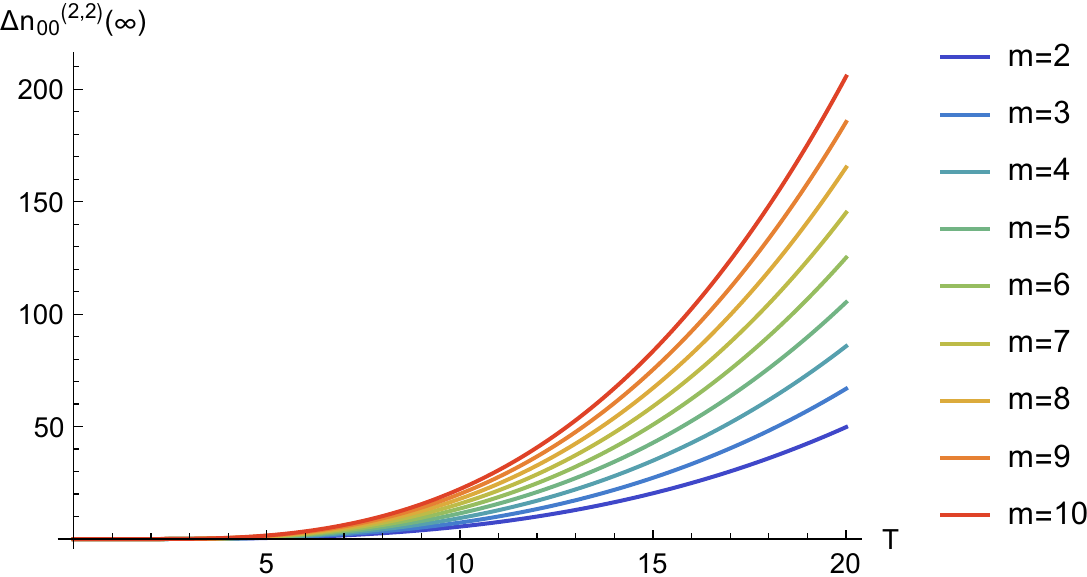}
    \includegraphics[width=0.49\textwidth]{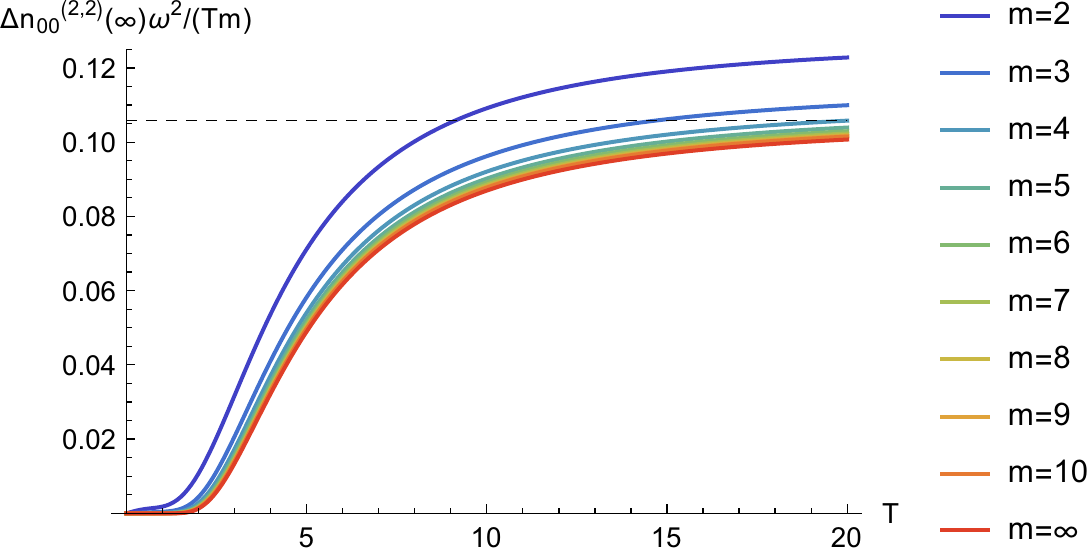}
    \caption{Prethermalization plateau of the momentum occupation
      {(in order $U^2E^2$)}
      with
      $\epsilon_\BK$ $=$ $\mybar{\epsilon}_\BK$ $=$ $0$ for the Hubbard model with
      diagonal field direction~\eqref{eq:H0gauss} subjected to the
      pump pulse~\eqref{eq:field-pulse}, unscaled (left) and scaled by
      pulse duration (right) for varying number of pulse oscillations
      $m$ and periods $T$. The limit of long pulse
      durations ($m$ $=$
      $\infty$) corresponds
      to~\eqref{eq:nTildeEQgauss}.
      Here the units are $\hbar/t^*$ for time,
        $t_*$ for energy, and $t^*/(ea)$
        for electric field, see
        remarks below~\eqref{eq:H0gauss}.
    }\label{fig:H0plateauScale2}
  \end{figure}

\begin{figure}[t]
    \centering
    \includegraphics[width=0.49\textwidth]{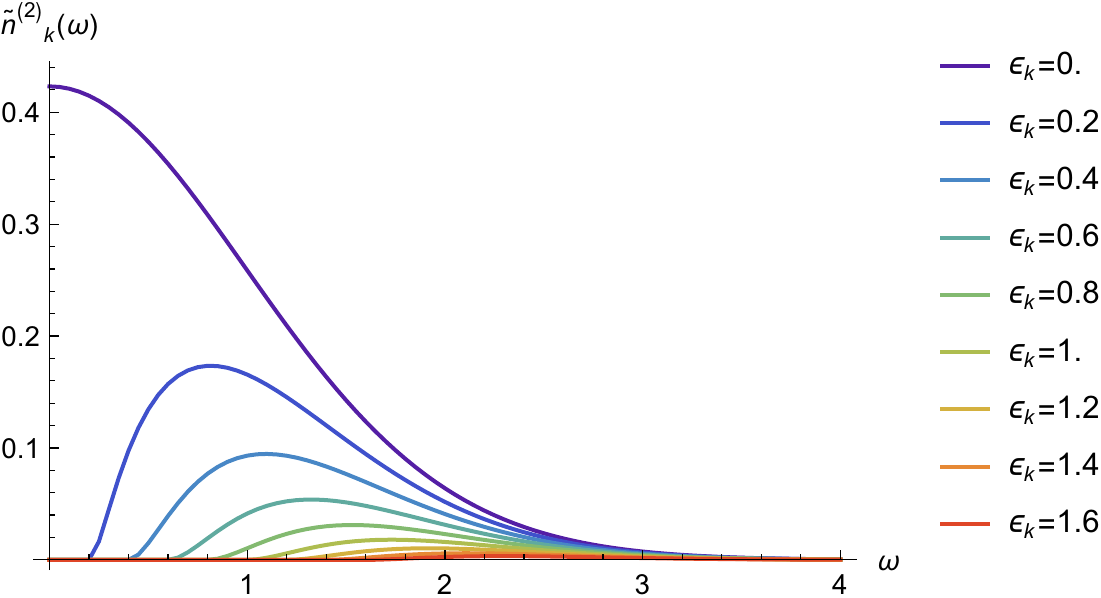}
    \includegraphics[width=0.49\textwidth]{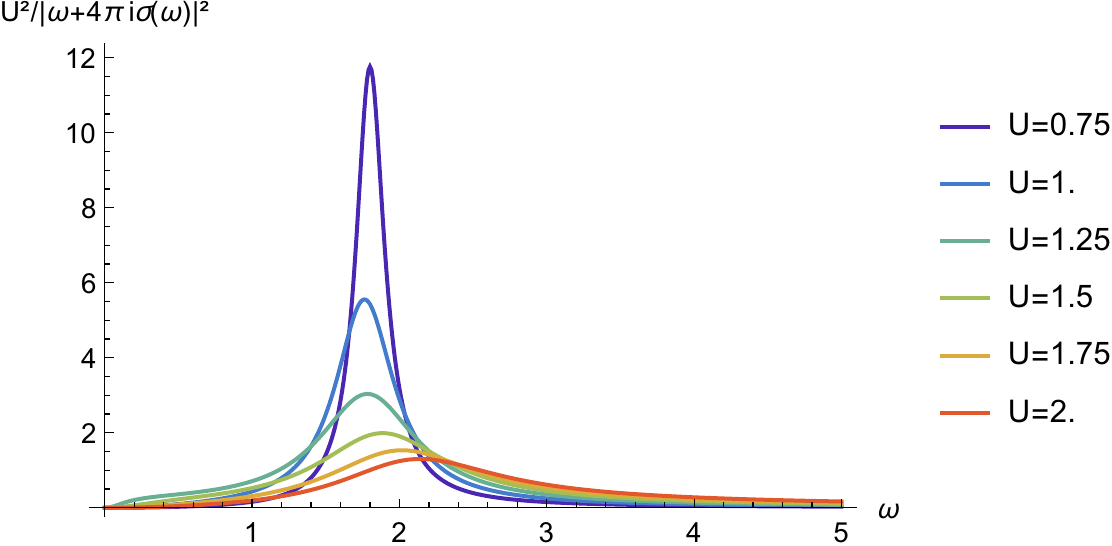}
    \caption{Contributions to the scaled prethermal plateau of
      $\Delta\langle\hat{n}_{\BK\sigma}\rangle_{t>\tFin}/\tFin$ in
      {both}
      second order in the electric field amplitude {and interaction} with
      $\mybar{\epsilon}_\BK$ $=$ $0$ and several $\epsilon_\BK$ for
      the Hubbard model with diagonal field
      direction~\eqref{eq:H0gauss} subjected to the pump
      pulse~\eqref{eq:field-pulse} in the limit of long pulse
      durations $\tFin$, as obtained from the $\BK$-dependent factor
      $\tilde{n}_{\BK}^{(2)}(\omega)$ of~\eqref{eq:nTildeEQ} (left)
      and the $U$-dependent factor in~\eqref{eq:nk-result}
      (right).
      Here the units are $t^*/\hbar$ for frequency,
        $e^2/(\hbar a)$ for $\sigma(\omega)$, $t_*$ for energy,
        $t^*/(ea)$ for electric field, and we have
        chosen $a\cdot|E_{\text{kin}}^{(0)}(0)|$ $=$ $4$ {\AA} $\cdot$ eV, see
        remarks below~\eqref{eq:H0gauss}.
    }\label{fig:nkLongTime}
  \end{figure}

  As the kinetic energy already thermalizes at the prethermal stage,
  its plateau corresponds to the energy absorbed from the pump
  pulse. Nevertheless the prethermal state differs from the eventual
  thermal state in its momentum occupation numbers of individual
  modes, as we now discuss. For them we also observe scaling behavior,
  as shown for one momentum in
  \textbf{Figure~\ref{fig:H0plateauScale2}},
  again with better scaling
  as the number of periods increases.  To obtain the leading term for
  the change in momentum occupation we use, similar
  to~\eqref{eq:absorptionExpansion},
  \begin{align}
    \Delta\langle\hat{n}_{\BK\sigma}\rangle_t
    &=
      \frac{-1}{c^2}
      \intlimits_{0}^{t}\!\D t_1
      \intlimits_{0}^{t_1}\!\D t_2
      \,A(t_1)A(t_2)
      \langle
      {e^2}
      \big[\hat{H}_0^{(1)}(t_2),\big[\hat{H}_0^{(1)}(t_1),\hat{n}_{\BK\sigma}(t)\big]\big]
      \rangle_{{{0}}}+O(A^3)
      \,.\label{eq:modeExpansion}
  \end{align}
  For a single mode we proceed {differently} than for the energy. We
  first obtain a result that holds independent of the interaction
  under the assumption that we may treat the Heisenberg operator
  $\hat{n}_{\BK\sigma}(t)$ as commuting with the Hamiltonian
  $\hat{H}$ in the long-time limit.  Then the dependence on $t_1$
  and $t_2$ in the above double commutator reduces to a time
  difference,
  \begin{align}
    \langle\big[\hat{H}_0^{(1)}(t_2),\big[\hat{H}_0^{(1)}(t_1),\hat{n}_{\BK\sigma}(t)\big]\big]\rangle_{{{0}}}
    \stackrel{~t\to\infty~}{=}
    \langle\big[\hat{H}_0^{(1)},\big[\hat{H}_0^{(1)}(t_1-t_2),\hat{n}_{\BK\sigma}(t)\big]\big]\rangle_{{{0}}}\,,
  \end{align}
  because the initial state is an eigenstate of the Hamiltonian
  $\hat{H}$ occurring in the Heisenberg propagators.  In terms of the
  Fourier components of the field pulse $E(t)$ $=$
  $\sum_j E_j \EXP^{\IMI t\omega_j}$ we find
  \begin{subequations}%
    \label{eq:prethermGeneral}
    \begin{align}%
      \lim\limits_{\tFin\to\infty}\frac{\Delta\langle\hat{n}_{\BK\sigma}\rangle_{t>\tFin}}{\tFin}
      &=
        \sum_j |E_j|^2\frac{\Tilde{n}_{\BK\sigma}(\omega_j)}{\omega_j^2}+O(E^3)\,,\label{eq:prethermGeneral1}\\
      \Tilde{n}_{\BK\sigma}(\omega)
      &=
        -
        \frac{{e^2}}{2}
        \lim\limits_{t\to\infty}
        \intlimits_{-\infty}^{\infty}\!\D\tau\,
        \EXP^{-\IMI\tau\omega}\langle\big[\hat{H}_0^{(1)},\big[\hat{H}_0^{(1)}(\tau),\hat{n}_{\BK\sigma}(t)\big]\big]\rangle_{{{0}}}\,.
        \label{eq:prethermGeneral2}
    \end{align}%
  \end{subequations}%
  which is still nonperturbative in the interaction. The
  steady-state momentum distribution difference thus scales linear
  with pulse duration, in analogy to
  \eqref{eq:heatingGeneral}
  for the absorbed energy, assuming that the limit
  in~\eqref{eq:prethermGeneral2} exists.
  It remains to evaluate it for weak interaction, for which we use
  the approach of Section~\ref{sec:pt}
  on~\eqref{eq:absorptionExpansion}. Its inner commutator may be
  written 
  \begin{align}
    \big[\hat{H}_0^{(1)}(t_1), \hat{n}_{\BK\sigma}(t) \big]
    &=
      \IMI g
      \intlimits_{-\infty}^{t_1}\!\D\tau_1\,f(\tau_1)
      \big[\big[\hat{V}_I(\tau_1),\hat{H}_0^{(1)}\big],
      \hat{n}_{\BK\sigma} \big]
      +
      \IMI g\intlimits_{-\infty}^{t}\!\D\tau\,f(\tau)
      \big[\hat{H}_0^{(1)}, \big[\hat{V}_I(\tau),\hat{n}_{\BK\sigma}
      \big]\big]+O(g^2)
      \nonumber\\
    &=\IMI g\intlimits_{t_1}^{t}\!\D\tau\, \big[\hat{H}_0^{(1)}, \big[\hat{V}_I(\tau),\hat{n}_{\BK\sigma} \big]\big]+O(g^2)\,,
  \end{align}
  by reordering the first commutator and combining the time
  integrations. The outer commutator then becomes
  \begin{align}
    \langle
    \big[\hat{H}_0^{(1)}(t_2),&\big[\hat{H}_0^{(1)}(t_1),\hat{n}_{\BK\sigma}(t)\big]\big]
                                \rangle_{{{0}}}
                                \nonumber\\&
    = \IMI g
    \intlimits_{-\infty}^{t_2}\!\D\tau_2\,f(\tau_2)\langle\big[\big[\hat{V}_I(\tau_2),\hat{H}_0^{(1)}\big],
    \big[\hat{H}_0^{(1)}(t_1),\hat{n}_{\BK\sigma}(t)\big]
    \big]\rangle_{(0)}+O(g^3)
    \nonumber\\
                              &= -g^2
                                \intlimits_{t_1}^{t}\!\D\tau\intlimits_{-\infty}^{t_2}\!\D\tau_2\,f(\tau_2)\langle\big[\big[\hat{V}_I(\tau_2),\hat{H}_0^{(1)}\big],\big[\hat{H}_0^{(1)},
                                \big[\hat{V}_I(\tau),\hat{n}_{\BK\sigma}
                                \big]\big]\big]\rangle_{(0)}+O(g^3)
                                \nonumber\\
                              &=
                                -g^2
                                \sum_l
                                (\EXP^{\IMI (t_1-t_2)\Delta E_l}
                                -
                                \EXP^{\IMI (t-t_2)\Delta E_l}
                                )\,|V_{0l}|^2\, \frac{(\Delta E^{(1)}_l)^2}{\Delta
                                E_l^2}\Delta n_{\BK,l}+\text{c.c.}
                                +O(g^3)\,,
  \end{align}
  where we inserted the noninteracting eigenstates in the last step,
  with $\Delta n_{\BK,l}$ $=$
  $\bra{\Psi_l}\hat{n}_{\BK\sigma}\ket{\Psi_l}$ $-$
  $\bra{\Psi_0}\hat{n}_{\BK\sigma}\ket{\Psi_0}$.  For a
  closely spaced band of energies $\Delta E_l$ we may assume that
  the second exponential factor drops out for large times $t$. To
  second order in the interaction we find
  \begin{subequations}%
    \begin{align}%
      \Tilde{n}_{\BK\sigma}(\omega)
      &=
        g^2\Tilde{n}^{(2)}_{\BK\sigma}(\omega)+O(g^3)
        \,,\\
      \Tilde{n}_{\BK\sigma}^{(2)}(\omega)
      &=\frac{{e^2}}{2}
        \intlimits_{-\infty}^{\infty}\!\D\tau\,
        \EXP^{-\IMI\tau\omega}
        \Big(\sum_l \frac{\EXP^{\IMI \tau\Delta E_l}}{\Delta E_l^2}|V_{0l}|^2
        (\Delta E^{(1)}_l)^2\Delta n_{\BK,l}+\text{c.c.} \Big)
        \,,
    \end{align}%
    \label{eq:nTildeEQ}%
  \end{subequations}%
  which we further evaluate for the Hubbard model with diagonal
  field direction,
  \begin{align}
    \Tilde{n}_{\BK\sigma}^{(2)}(\omega)
    &=\Theta(|\omega|-\epsilon_{\BK\sigma})
      \,
      \Big(\frac{3}{2}+\mybar{\epsilon}_\BK^2\Big)
      \intlimits_{\frac{1}{3}}^{1}\!\D b\,a_3(b)
      \,
      \Big(1-\frac{\epsilon_\BK}{|\omega|} \Big)^2
      \,
      \frac{\EXP^{-b\omega^2}}{\sqrt{\pi}}
      \label{eq:nTildeEQgauss}\,.
  \end{align}
  Finally we insert the enveloped field pulse~\eqref{eq:field-pulse}
  with pump frequency $\omega_\text{pump}$ $=$ $2\pi/T$, yielding
  \begin{align}
    \lim\limits_{m\to\infty}\frac{\Delta\langle\hat{n}_{\BK\sigma}\rangle_{t>mT}}{mT}
    &=
      \frac{E^2 U^2\Tilde{n}^{(2)}_{\BK\sigma}(\omega_\text{pump})}{4\omega_\text{pump}^2}
      +O(U^3E^2)+O(U^2E^3)
      \nonumber\\
    &=
      \frac{E_{\text{ext}}^2 U^2\Tilde{n}^{(2)}_{\BK\sigma}(\omega_\text{pump})}{4|\omega_\text{pump}+4\pi \IMI \sigma(\omega_\text{pump})|^2}
      +O(U^3E^2)+O(U^2E^3)\,.\label{eq:nk-result}
  \end{align}
  This prethermal state is plotted in Figure~\ref{fig:H0plateauScale2}
  based on~\eqref{eq:nTildeEQgauss}, showing that the scaling is well
  attained already for rather small $m$.  In general, two factors
  contribute to the prethermalization plateau of the momentum
  occupation numbers, which are shown in
  \textbf{Figure~\ref{fig:nkLongTime}}%
  {with parameters as given below~\eqref{eq:H0gauss}}. While the precise response
  depends on the momentum, the strongest effect will always occur for
  pump frequencies near the resonance ($\omega_*$ $\simeq$ $1.88$).
  Since the resonance peak becomes larger for small $U$, we conclude
  that long-lived prethermalization plateaus can be excited by
  electric fields.  Inversely, the value of $U$ could be estimated in
  principle by locating the resonance frequency for a known band
  structure. The result~\eqref{eq:nk-result} corresponds
  to~\eqref{eq:nk-intro-result} in the introduction.  There, the
  prefactor $s$ $=$ $1/4$ 
  was split off, which depends on the shape of the envelope. For
  example, an increase to the maximum amplitude $E$ of the pulse that
  is steeper than in~\eqref{eq:field-pulse} would result in a larger
  $s$, which nevertheless remains on the order of unity.

  \subsection{Prethermal vs. thermal steady states}
  \label{sec:ppcond}
  
  The prethermalization plateaus~\eqref{eq:nk-result} for the
  momentum occupation numbers are proportional to $E^2U^2$, i.e.,
  they occur in second order in the field \emph{and} interaction,
  while in lower orders the momentum occupation always relaxes back
  to its initial distribution after the pulse. The prethermal value
  of $n_{\BK\sigma}$ could be observed on the one hand by a
  momentum-resolved probe, such as time-resolved angle-resolved
  photoemission spectroscopy, as individual momentum occupation
  numbers, especially close to the Fermi surface, relax rather
  slowly to their prethermal value and will subsequently exhibit
  further relaxation to the thermal state.  Using optical
  spectroscopy, on the other hand, it is more difficult to
  distinguish the prethermalization plateau from the thermal state,
  i.e., through the time dependence of the conductivity as derived
  below. This difficulty stems from the fast relaxation of the
  kinetic energy on the prethermalization time scale $1/t_*$, as
  seen in Figure~\ref{fig:ObsVara}, and it also undergoes no further
  relaxation since it has then already attained its thermal value in
  order $U^2$. To understand this more quantitatively, we consider
  the conductivity $\sigma^{\text{pp}}(t,\tau)$ for a probe pulse in
  linear response when the system is subjected to the pump pulse.
  For this `pump-probe conductivity' we thus have
  \begin{subequations}%
    \begin{align}%
      \sigma^{\text{pp}}(t,\tau)
      &=
        \sigma^{\text{probe,dia}}(t)+\sigma^{\text{probe,pm}}(t)
        \,,\\
      \sigma^{\text{pp,dia}}(t)
      &=
        {e^2}
        \,
        \langle\hat{H}_0^{(2)}\rangle_{t}
        \,,~~~~
        \sigma^{\text{pp,pm}}(t,\tau)
        =
        {e^2}
        \,
        \IMI\intlimits_{\tau}^{t}\!\D\tau'\,
        \langle\big[\hat{H}_0^{(1)}(\tau'),\hat{H}_0^{(1)}(t)\big]\rangle_0
        \,.
    \end{align}%
    \label{eq:ppCurrentResponse}%
  \end{subequations}%
  Here the Heisenberg operators evolve with the operator
  $\hat{H}+\delta \hat{H}_0(t)$ which describes the pump pulse as
  given given earlier.  After the pulse the diamagnetic contribution
  $\sigma^{\text{pp}}(t,\tau)$ relaxes similar to the kinetic energy
  as described above, so that it remains to evaluate the
  paramagnetic contribution
  using~\eqref{eq:PulseInteraction}, 
  \begin{align}
    \langle\big[\hat{H}_0^{(1)}(\tau'),\hat{H}_0^{(1)}(t)\big]\rangle_{{{0}}}
    &=
      g^2
      \sum_l |V_{0l}|^2 \big(E_l^{(1)}\big)^2
      \intlimits_{-\infty}^{t}\!\D t_1\intlimits_{-\infty}^{\tau'}\!\D\tau_1\,
      f(t_1)f(\tau_1)
      \EXP^{\IMI\intlimits_{\tau_1}^{t_1}\!\D t'\,(\Delta E_{l}+\delta
      e_{l}(t'))}
      -\text{c.c.} +O(g^3)
      \nonumber\\
    &=
      g^2
      \sum_l |V_{0l}|^2 \big(E_l^{(1)}\big)^2
      \EXP^{\IMI (t-\tau')\Delta E_l}\varphi_l^{(1)}(t)^* \varphi_l^{(1)}(\tau')
      -\text{c.c.} +O(g^3)
      \,.
  \end{align}
  We consider  $t$ $>$ $\tFin$  and  for simplicity assume that pump and probe pulse do not overlap,
  so that  only $\tau$ $>$ $\tFin$ contributes and  the upper limit of the integrals can be replaced by
  $\tFin$, 
  \begin{align}
    &\varphi_l^{(1)}(t)
      =
      \frac{-1}{\Delta E_l}+\IMI\EXP^{\IMI t \Delta E_l} h_l
      \,,~~~~
      h_l
      =
      -\intlimits_{0}^{\tFin}\!\D t_1\,
      \frac{\EXP^{-\IMI t_1\Delta E_l}}{\Delta E_l}\EXP^{\IMI
      \intlimits_{t_1}^{\tFin}\!\D t' \delta e_l(t')
      }\delta e_l(t_1)
      \,,
    \\
    &\RE\,
      \varphi_l^{(1)}(t)^*\,\IMI\intlimits_{\tau}^{t}\!\D\tau'\EXP^{\IMI
      (t-\tau')\Delta E_l} \varphi_l^{(1)}(\tau')
    =\RE\,
    \frac{\EXP^{\IMI (t-\tau)\Delta E_l}-1}{\Delta
    E_l^3}+\frac{\EXP^{\IMI (t-\tau)\Delta E_l}-1}{\Delta E_l^2}
    \EXP^{-\IMI t \Delta E_l}i h_l^*+ \frac{t-\tau}{\Delta E_l}
    \EXP^{\IMI t \Delta E_l} h_l
    \,.
  \end{align}
  Here the first term gives the paramagnetic conductivity in
  equilibrium, while the other terms lead to the transient part of the
  paramagnetic pump-probe conductivity, which we denote by
  \begin{align}
    \Delta\sigma^{\text{pp,pm},(2)}(t,\tau)
    =
    {e^2}
    \,&
              \intlimits_{0}^{\tFin}\!\D t_1
              \sum_l |V_{0l}|^2\frac{\big( \Delta E_l^{(1)}\big)^2}{\Delta E_l^2}
              \nonumber\\&~~~~~~\times
    \EXP^{\IMI \intlimits_{t_1}^{\tFin}\!\D t' \delta e_l(t')
    }\delta e_l(t_1)\Big(\IMI\frac{\EXP^{\IMI \tau \Delta
    E_l}-\EXP^{\IMI t \Delta E_l}}{\Delta E_l}-\EXP^{\IMI t\Delta
    E_l}(t-\tau)\Big)+\text{c.c.}
    \,.
  \end{align}
  Since each term contains a factor $\EXP^{\IMI t \Delta E_l}$ or
  $\EXP^{\IMI \tau \Delta E_l}$ this contribution is suppressed for $t$ $>$
  $\tau\to\infty$ when integrated over the band energies $E_l$. For
  examples, for the Hubbard model with diagonal field direction
  $\Delta\sigma^{\text{pp,pm},(2)}(t,\tau)\to0$ vanishes proportional
  to a Gaussian $\sim\EXP^{-\tau^2/4}$. As a consequence, the
  pump-probe conductivity as an integrated quantity is not well-suited
  to observe the prethermal state, as discussed above.

  \section{Conclusion}
  \label{sec:conclusion}

  Prethermalized states can in general be generated by time-dependent
  switching or driving protocols of the interaction or an external
  field. Here we focused on enveloped electric field pulses which add
  a time-dependent modulation to the kinetic energy. After the pulse
  the electronic system relaxes to a prethermal steady state on short
  timescales~$\sim$ $\hbar/bandwidth$. Even for strong fields this
  behavior is well-described in the leading quadratic orders of
  interaction and field strength. The response to the pump field will
  be enhanced for pump pulses near an interaction-dependent resonance
  frequency which develops due to the field response inside the
  sample.
  {On the other hand, the details of the pulse shape are found
    to be less important, as they are merely enter the prefactor $s$
    in the leading-order result~\eqref{eq:nk-intro-result}.}
  From an analysis of the real-time conductivity we concluded
  that momentum-resolved probe techniques are typically necessary to
  distinguish the prethermal from the thermal state. Our explicit
  evaluations were performed for a Hubbard model with diagonal field
  direction in high dimensions, but could be extended to other
  Hubbard-type systems. For example, the effect of features in the
  band dispersion or of band degeneracies would be of particular
  interest. Our general perturbative approach may also be useful to
  provide input into effective models for later relaxation stages as
  well as in other contexts.

  \medskip
  \noindent\textbf{Acknowledgments} \par

  \noindent This work was supported in part by Deut\-sche
  For\-schungs\-ge\-mein\-schaft under project
  number 
  107745057 (TRR 80).


  \appendix
  \section*{Appendix}
  
  In this appendix we provide technical details of the evaluations for the 
  Hubbard model with diagonal field
  direction~\eqref{eq:H0gauss} for the perturbative approach of Section~\ref{sec:pt}. The vector potential effectively
  enters in time arguments of the form 
  $\tau_1$ $=$ $\int\D\tau\,\cos A(\tau)$ and $\tau_2$ $=$
  $\int\D\tau\,\sin A(\tau)$, so that we need to evaluate the following expectation
  values in the noninteracting ground state $|\Psi_0\rangle$
  \begin{align}
    \langle \big[ \hat{D},
    \big[\hat{D}_I(\tau_1,\tau_2),\hat{n}_{\BK\sigma}\big] \big]
    \rangle_{(0)}
    &=
      2\,\RE
      \sum_l |D_{0l}|^2
      \Delta n_{\BK,l}
      \EXP^{\IMI \tau_1 \Delta E_l+\IMI\tau_2 \Delta \mybar{E}_m}
      \nonumber\\&
    =
    2\,\RE
    \sum_{i,\lambda=\pm}\lambda
    f^{\lambda}_{ \BK\sigma}(\bm{R}_i,\tau_1,\tau_2)
    F^{-\lambda}_{\sigma}(\bm{R}_i,\tau_1,\tau_2)
    \prod\limits_{\lambda'=\pm}F^{\lambda'}_{\mybar{\sigma}}(\bm{R}_i,\tau_1,\tau_2)
    \,,\\
    \IMI\langle \big[\hat{D},\hat{D}_I(\tau_1,\tau_2) \big] \rangle_{(0)}
    &=
      2\,\RE
      \sum_l |D_{0l}|^2
      \,\IMI\,\EXP^{\IMI \tau_1 \Delta E_l+\IMI\tau_2 \Delta
      \mybar{E}_m}
      \nonumber\\&
    =
    \,\RE\,\IMI
    \sum_i
    \prod\limits_{\lambda=\pm}F^{\lambda}_{\mybar{\sigma}}(\bm{R}_i,\tau_1,\tau_2)F^{\lambda}_{\sigma}(\bm{R}_i,\tau_1,\tau_2)
    \,,
  \end{align}
  in terms of $D_{nm}$ $=$
  $\langle\Psi_n|\N{i\uparrow}\N{i\downarrow}|\Psi_m\rangle$,
  $\delta\hat{\mybar{H}}_0\ket{\Psi_l}$ $=$ $\mybar{E}_l\ket{\Psi_l}$
  $\Delta\mybar{E}_l$ $=$ $\mybar{E}_l-\mybar{E}_0$, and the functions
  ($\lambda$ $=$ $\pm$)
  \begin{align}
    f^\lambda_{\BK\sigma}(\bm{R}_i,\tau_1,\tau_2)
    =
    \EXP^{-\IMI(\bm{R}_i\cdot\BK+\tau_1\epsilon_{\BK}+\tau_2\mybar{\epsilon}_{\BK})}
    \langle \delta_{\lambda+}+\lambda\hat{n}_{\BK\sigma}\rangle_{(0)}
    \,,~~~~~~
    F^{\lambda}_{\sigma}(\bm{R}_i,\tau_1,\tau_2)
    =
    \frac{1}{L}\sum_{\BK} f^{\lambda}_{\BK\sigma}(\bm{R}_i,\tau_1,\tau_2)\,.
  \end{align}
  In the limit of high dimensions only the local term $\bm{R}_i$ $=$
  $\bm{0}$ contributes, leading to integrals over the density of
  states~\eqref{eq:H0gauss3},
  \begin{align}
    F^{\lambda}_{\sigma}(\tau_1,\tau_2)
    &=
      \int\D\epsilon\int\D\mybar{\epsilon}\,\rho(\epsilon,\mybar{\epsilon})\,\EXP^{-\lambda\IMI(\tau_1 \epsilon+\tau_2 \mybar{\epsilon})}
      \langle \delta_{\lambda+}+\lambda\hat{n}_{\epsilon \mybar{\epsilon}\sigma}\rangle_{(0)}
      \nonumber\\
    =
    F(\tau_1,\tau_2)
    &=\intlimits_{0}^{\infty}\D\epsilon
      \frac{\EXP^{-\epsilon^2+\IMI\epsilon \tau_1}}{\sqrt{\pi}}
      \,
      \intlimits_{-\infty}^{\infty}\D\mybar{\epsilon}
      \frac{\EXP^{-\mybar{\epsilon}^2+\IMI\mybar{\epsilon} \tau_2}}{\sqrt{\pi}}    
      \,,
  \end{align}
  where the second line, independent of $\lambda$ and $\sigma$, applies
  for the paramagnetic ground state of the half-filled band.
  For the expectation values
  \begin{align}
    D^{(1)}(t) = \sum_l  |D_{0l}|^2 \varphi_l^{(1)}(t)+\text{c.c.}\,,~~~~
    n_{\BK\sigma}^{(2)}(t) = \sum_l  |D_{0l}|^2 \Delta n_{\BK,l} \varphi_l^{(2)}(t)\,,
  \end{align}
  we require $\varphi^{(1)}_l(t)$ and $\varphi^{(2)}_l(t)$
  from~\eqref{eq:extractPhi1} and~\eqref{eq:extractPhi2}, which
  involve $\delta e_l(t_2)$ $=$
  $\Delta E_l\,(\cos A(t_2)-1)+\Delta\mybar{E}_l\,\sin A(t_2)$.
  We are left with powers of $\Delta E_l$ and $\Delta\mybar{E}_l$,
  which we rewrite as
  \begin{align}
    (\Delta E_l)^{n_1} (\Delta \mybar{E}_l)^{n_2}\EXP^{\IMI \tau_1 \Delta E_l+\IMI\tau_2 \Delta \mybar{E}_l}
    =
    \Big(\frac{\partial}{\partial\,\IMI \tau_1}\Big)^{n_1}\,
    \Big(\frac{\partial}{\partial\,\IMI \tau_2}\Big)^{n_2}\,
    \EXP^{\IMI \tau_1 \Delta E_l+\IMI\tau_2 \Delta \mybar{E}_l}
    \,,\label{eq:integration}
  \end{align}
  with exponents $n_1\in \{-2,-1,0,1\}$ and $n_2\in
  \{0,1,2\}$. Then the sum over states can be performed,
  resulting again integrals $F(\tau_1,\tau_2)$ over the density
  of states. Its differentiation is
  straightforward, while for positive exponents the
  integrations (with appropriate integration
  limits) are best simplified by another
  transformation. Consider first the functions appearing in
  $D^{(1)}(t)$, 
  \begin{align}
    F(\tau_1,\tau_2)^m
    &=
      \bigg( \intlimits_{-\infty}^{\infty}\D\mybar{\epsilon} \frac{\EXP^{-\mybar{\epsilon}^2+\IMI\mybar{\epsilon} \tau_2}}{\sqrt{\pi}} \bigg)^m
      \intlimits_{0}^{\infty}\D\epsilon_1 \ldots \intlimits_{0}^{\infty}\D\epsilon_m \frac{\EXP^{-\sum_{l=1}^{m}\epsilon_l^2+\IMI\epsilon_l \tau_1}}{\sqrt{\pi^m}}
    \\
    &=
      \EXP^{-\frac{m\tau_2^2}{4} } \intlimits_{0}^{\infty}\D\epsilon_0\, \frac{\EXP^{\IMI \tau_1 \epsilon_0}}{\sqrt{\pi^m}} \intlimits_{0}^{\infty}\D\epsilon_1 \ldots \intlimits_{0}^{\infty}\D\epsilon_m\, \delta(\epsilon_0-\sum_{l=1}^{m}\epsilon_l) \EXP^{-\sum_{l=1}^{m}\epsilon_l^2}
    \,,
  \end{align}
  which can be expressed in terms of a weight function
  \begin{align}
    a_n(b)&=\intlimits_0^{\infty}\D x_1 \ldots \intlimits_0^{\infty}\D x_n\, \delta(1-\sum_{i=1}^n x_i)\delta(b-\sum_{j=1}^n x_j^2)\,.\label{eq:an}
  \end{align}
  Integrating $m-1$ times we obtain
  \begin{align}
    \Big(\frac{\partial}{\partial\,\IMI \tau_1}\Big)^{1-m}\,
    F(\tau_1,\tau_2)^m
    &=
      \EXP^{-\frac{m\tau_2^2}{4} } \intlimits_{0}^{\infty}\D b
      \intlimits_{0}^{\infty}\D\epsilon_0\,
      \frac{\EXP^{\IMI \tau_1\epsilon_0-b\epsilon_0^2}}{\sqrt{\pi^m}} a_m(b)
      \,,
    \\
    \Big(\frac{\partial}{\partial\,\IMI \tau_1}\Big)^{1-m}\, F(\tau_1,\tau_2)^m
    &=
      \EXP^{-\frac{m\tau_2^2}{4} } \intlimits_{\frac{1}{m}}^{1}\D b\,
      \frac{a_m(b)}{\sqrt{b \pi^m}}
      \Big(\frac{\sqrt{\pi}}{2}\EXP^{-\frac{\tau_1^2}{4b}}+\IMI\,\DawsonF(\tfrac{\tau_1}{2\sqrt{b}})\Big)
      \,,
  \end{align}
  where the Dawson function $\DawsonF(x)$ was defined
  in~\eqref{eq:conductivityPM}.  We need only the cases $n$ $=$
  $3,4$, for which~\cite{alexander_notitle_nodate}
  \begin{align}
    a_3(b)&=
            \begin{cases}
              \dfrac{\pi}{\sqrt{3}}
              &\text{if~}\frac13\leq b\leq\frac12,
              \\[1ex]
              \dfrac{\pi}{\sqrt{3}}
              -
              \sqrt{3}\arccos\dfrac{1}{\sqrt{6b-2}}
              &\text{if~}\frac12\leq b\leq1,
              \\[1ex]
              0&\text{otherwise,}
            \end{cases}
                 \label{eq:a3}
    \\
    a_4(b)&=
            \begin{cases}
              \pi \sqrt{b-\frac{1}{4}}
              &
              \text{if~} \frac{1}{4} < b < \frac{1}{3},
              \\
              \dfrac{\pi}{\sqrt{3}}-\pi \sqrt{b-\frac{1}{4}}
              &
              \text{if~} \frac{1}{3} < b < \frac{1}{2},
              \\
              \sqrt{3}\,\arcsin\dfrac{1}{\sqrt{6b-2}}+3\sqrt{b-\frac{1}{4}}\,\arcsin\dfrac{\sqrt{1-6b+8b^2}}{3b-1}-\dfrac{\pi}{2\sqrt{3}}-\pi \sqrt{b-\frac{1}{4}}
              & \text{if~} \frac{1}{2} < b < 1,
              \\
              0 & \text{otherwise.}
            \end{cases}
                  \!\!\!\!\!\!\label{eq:a4}
  \end{align}
  Integrations are thus avoided for ${m-1+n_1\geq 0}$ with
  $m\in\{3,4\}$, except for one final numerical integration over
  $b$. For $n_{\BK\sigma}^{(2)}(t)$, on the other hand, an integration
  over $\epsilon_0$ also remains. In this case we need (for
  $\epsilon>0$)
  \begin{align}
    \EXP^{\IMI \tau_1 \epsilon} F(\tau_1,\tau_2)^3
    &=
      \EXP^{-\frac{3\tau_2^2}{4}}
      \intlimits_{\epsilon}^{\infty}\D\epsilon_0\,
      \frac{\EXP^{\IMI \tau_1 \epsilon}}{\sqrt{\pi^3}}
      \intlimits_{0}^{\infty}\D\epsilon_1
      \intlimits_{0}^{\infty}\D\epsilon_2
      \intlimits_{0}^{\infty}\D\epsilon_3\,
      \delta(\epsilon_0 -\epsilon-\epsilon_1-\epsilon_2-\epsilon_3)
      \EXP^{-\epsilon_1^2-\epsilon_2^2-\epsilon_3^2}
      \nonumber\\
    &=
      \EXP^{-\frac{3\tau_2^2}{4}}
      \intlimits_{\epsilon}^{\infty}\D\epsilon_0\,
      \epsilon_0^2
      \intlimits_{0}^{\infty}\D b\,
      \frac{\EXP^{\IMI \tau_1 \epsilon-b\epsilon_0^2}}{\sqrt{\pi^3}}\,
      a_3\Big(\frac{b\epsilon_0^2}{(\epsilon_0-\epsilon)^2}\Big)\,.
  \end{align}
  Integrating twice and rearranging the integrals we arrive at 
  \begin{align}
    \Big(\frac{\partial}{\partial\,\IMI \tau_1}\Big)^{-2}\,
    \EXP^{\IMI \tau_1\epsilon}F(\tau_1,\tau_2)^3
    =
    \EXP^{-\frac{3\tau_2^2}{4} } \intlimits_{\frac{1}{3}}^{1}\D b\,
    a_3(b)
    \intlimits_{\epsilon}^{\infty}\frac{\D\epsilon_0}{\epsilon_0^2}\,
    (\epsilon_0-\epsilon)^2\,
    \frac{\EXP^{\IMI \tau_1 \epsilon_0-b\epsilon_0^2}}{\sqrt{\pi^3}}
    \,,\label{eq:a3Trick}
  \end{align}
  which can be differentiated analytically with respect to $\tau_1$
  and $\tau_2$ as needed, leaving numerical integrations over $b$,
  $t_1$, and $t_2$.  In the limit $\epsilon \to 0$, we recover the
  previous result.  For $\epsilon$ $>$ $0$ we integrate numerically
  over $\epsilon_0$ as well.

\end{document}